\documentclass[sigconf]{acmart}
\usepackage[vlined,boxed]{algorithm2e}
\usepackage{amsmath}
\usepackage{amsfonts}
\usepackage{epsfig}
\usepackage{graphics}
\usepackage{amssymb}
\usepackage{color}
\usepackage{comment}
\usepackage{paralist}
\usepackage{mathtools}
\usepackage{multirow}
\usepackage{epstopdf}
\usepackage{subfigure}
\usepackage{eepic}
\usepackage{stmaryrd}
\usepackage{textcomp}
\usepackage{balance}
\usepackage[inline]{enumitem}
\usepackage{ulem}

\renewcommand{\baselinestretch}{.97}

\begin{document}

\newcommand{\limpl}{\ensuremath{\rightarrow}} 
\newcommand{\rimpl}{\ensuremath{\leftarrow}}
\newcommand{\fk}[1]{\ensuremath{\mathbf{#1}}} 
\newcommand{\ve}[1]{\ensuremath{\bar{#1}}} 
\newcommand{\ca}[1]{\ensuremath{\mathcal{#1}}} 
\newcommand{\tup}[1]{\ensuremath{(#1)}} 
\newcommand{\set}[1]{\ensuremath{\{#1\}}} 

\newcommand{\size}[1]{\ensuremath{|#1|}} 
\newcommand{\sche}[1]{\ensuremath{\mathbf{#1}}}
\newcommand{\ata}[1]{\ensuremath{\mathcal{#1}}}

\newcommand{\nchase}[3]{\ensuremath{\mathrm{chase}^{#1}(#2,#3)}}
\newcommand{\Ra}{\ensuremath{\Rightarrow}}
\newcommand{\La}{\ensuremath{\Leftarrow}}
\newcommand{\rk}[1]{\ensuremath{\mathrm{rk}(#1)}}
\newcommand{\pred}[1]{\ensuremath{\mathrm{succ}(#1)}}
\newcommand{\preds}[3]{\ensuremath{\mathrm{succ}_{#2,#3}(#1)}}
\newcommand{\succs}[3]{\ensuremath{\mathrm{succ}_{#2,#3}(#1)}}
\newcommand{\predi}[2]{\ensuremath{\mathrm{succ}^{#2}(#1)}}
\newcommand{\predss}[4]{\ensuremath{\mathrm{succ}^{#2}_{#3,#4}(#1)}}
\newcommand{\dpth}[1]{\ensuremath{\mathrm{dp}(#1)}}
\newcommand{\lvl}[1]{\ensuremath{\ell(#1)}}
\newcommand{\lvli}[2]{\ensuremath{\ell_{#2}(#1)}}
\newcommand{\rlvl}[2]{\ensuremath{\mathrm{level}_{#2}(#1)}}
\newcommand{\rec}[1]{\ensuremath{\mathsf{rec}(#1)}}
\newcommand{\ldpth}[1]{\ensuremath{\mathrm{ldp}(#1)}}
\newcommand{\last}[1]{\ensuremath{\mathrm{last}(#1)}}
\newcommand{\jg}[1]{\ensuremath{\mathbb{J}(#1)}}
\newcommand{\stratumi}[2]{\ensuremath{s_{#2}(#1)}}

\newcommand{\TODO}[1]{\par\noindent \textcolor{red}{\textbf{TODO:} #1}}

\newcommand{\mi}[1]{\mathit{#1}}
\newcommand{\ins}[1]{\mathbf{#1}}
\newcommand{\cali}[1]{\mathcal{#1}}
\newcommand{\adom}[1]{\mathsf{dom}(#1)}
\renewcommand{\paragraph}[1]{\textbf{#1}}
\newcommand{\ra}{\rightarrow}
\newcommand{\la}{\leftarrow}
\newcommand{\fr}[1]{\mathsf{front}(#1)}
\newcommand{\dep}{\Sigma}
\newcommand{\sch}[1]{\mathsf{sch}(#1)}
\newcommand{\edb}[1]{\mathsf{edb}(#1)}
\newcommand{\body}[1]{\mathsf{body}(#1)}
\newcommand{\head}[1]{\mathsf{head}(#1)}
\newcommand{\guard}[1]{\mathsf{guard}(#1)}
\newcommand{\ward}[1]{\mathsf{ward}(#1)}
\newcommand{\class}[1]{\mathsf{#1}}
\newcommand{\pos}[1]{\mathsf{pos}(#1)}
\newcommand{\app}[2]{\langle #1,#2 \rangle}
\newcommand{\crel}[1]{\prec_{#1}}
\newcommand{\base}[1]{\mathsf{base}(#1)}
\newcommand{\mods}[2]{\mathsf{mods}(#1,#2)}
\newcommand{\cert}[3]{\mathsf{cert}(#1,#2,#3)}
\newcommand{\qans}{{\sf CQAns}}
\newcommand{\eval}{{\sf Eval}}
\newcommand{\chase}[2]{\mathsf{chase}(#1,#2)}
\newcommand{\aff}[1]{\mathsf{aff}(#1)}
\newcommand{\nonaff}[1]{\mathsf{nonaff}(#1)}
\newcommand{\pg}[1]{\mathsf{pg}(#1)}
\newcommand{\pnf}[1]{\mathsf{pnf}(#1)}
\newcommand{\var}[1]{\mathsf{var}(#1)}
\newcommand{\atoms}[1]{\mathsf{atoms}(#1)}
\newcommand{\eq}{{\sf eq}}
\newcommand{\comp}[1]{\mathsf{comp}(#1)}
\newcommand{\ido}{\mi{ido}}
\newcommand{\wid}[1]{\mathsf{wd}(#1)}
\newcommand{\nwd}[1]{\mathsf{nwd}(#1)}
\newcommand{\lwd}[1]{\mathsf{lwd}(#1)}
\newcommand{\pol}{\mathsf{pol}}
\newcommand{\N}[1]{\mathsf{n}(#1)}
\newcommand{\corr}{\mathsf{UnboundedTiling}}
\newcommand{\ex}[3]{\mathsf{exp}_{#1}^{#2}(#3)}
\newcommand{\ep}[1]{\mathsf{ep}(#1)}
\newcommand{\cep}[1]{\mathsf{cep}(#1)}
\newcommand{\pep}[1]{\mathsf{pep}(#1)}


\def\qed{\hfill{\qedboxempty}      
  \ifdim\lastskip<\medskipamount \removelastskip\penalty55\medskip\fi}

\def\qedboxempty{\vbox{\hrule\hbox{\vrule\kern3pt
                 \vbox{\kern3pt\kern3pt}\kern3pt\vrule}\hrule}}

\def\qedfull{\hfill{\qedboxfull}   
  \ifdim\lastskip<\medskipamount \removelastskip\penalty55\medskip\fi}

\def\qedboxfull{\vrule height 4pt width 4pt depth 0pt}

\newcommand{\markfull}{\qedboxfull}
\newcommand{\markempty}{\qed}


\newcommand{\OMIT}[1]{}

\newtheorem{claim}[theorem]{Claim}
\newtheorem{fact}[theorem]{Fact}
\newtheorem{observation}{Observation}
\newtheorem{apptheorem}{Theorem}[section]
\newtheorem{appcorollary}[apptheorem]{Corollary}
\newtheorem{appproposition}[apptheorem]{Proposition}
\newtheorem{applemma}[apptheorem]{Lemma}
\newtheorem{appclaim}[apptheorem]{Claim}
\newtheorem{appfact}[apptheorem]{Fact}

\fancyhead{}

\setcopyright{none}
\settopmatter{printacmref=false} 
\renewcommand\footnotetextcopyrightpermission[1]{} 
\pagestyle{plain} 

\title{The Space-Efficient Core of Vadalog}

\author{Gerald Berger}
\affiliation{%
	\institution{Institute of Logic and Computation}
	\city{TU Wien}
}

\author{Georg Gottlob}
\affiliation{%
	\institution{Department of Computer Science}
	\city{University of Oxford \& TU Wien}
}

\author{Andreas Pieris}
\affiliation{%
	\institution{School of Informatics}
	\city{University of Edinburgh}
}

\author{Emanuel Sallinger}
\affiliation{%
	\institution{Department of Computer Science}
	\city{University of Oxford}
}

\begin{abstract}
Vadalog is a system for performing complex reasoning tasks such as those required in advanced knowledge graphs. The logical core of the underlying Vadalog language is the warded fragment of tuple-generating dependencies (TGDs). This formalism ensures tractable reasoning in data complexity, while a recent analysis focusing on a practical implementation led to the reasoning algorithm around which the Vadalog system is built. A fundamental question that has emerged in the context of Vadalog is the following: can we limit the recursion allowed by wardedness in order to obtain a formalism that provides a convenient syntax for expressing useful recursive statements, and at the same time achieves space-efficiency?
After analyzing several real-life examples of warded sets of TGDs provided by our industrial partners, as well as recent benchmarks, we observed that recursion is often used in a restricted way: the body of a TGD contains at most one atom whose predicate is mutually recursive with a predicate in the head. We show that this type of recursion, known as piece-wise linear in the Datalog literature, is the answer to our main question.
We further show that piece-wise linear recursion alone, without the
wardedness condition, is not enough as it leads to the undecidability
of reasoning.
We finally study the relative expressiveness of the query languages based on (piece-wise linear) warded sets of TGDs.
\end{abstract}

\maketitle

\section{Introduction}\label{sec:introduction}

In recent times, thousands of companies world-wide wish to manage their own knowledge graphs (KGs), and are looking for adequate knowledge graph management systems (KGMS).
The term knowledge graph originally only referred to Google's Knowledge Graph, i.e., ``a knowledge base used by Google and its services to enhance its search engine's results with information gathered from a variety of sources.\footnote{https://en.wikipedia.org/wiki/Knowledge\_Graph}'' In the meantime, several other large companies have constructed their own knowledge graphs, and many more companies would like to maintain a private corporate knowledge graph incorporating large amounts of data in form of database facts, both from corporate and public sources, as well as rule-based knowledge. Such a corporate knowledge graph is expected to contain relevant business knowledge, for example, knowledge about customers, products, prices, and competitors, rather than general knowledge from Wikipedia and similar sources. It should be managed by a KGMS, that is, a knowledge base management system, which performs complex rule-based reasoning tasks over very large amounts of data and, in addition, provides methods and tools for data analytics and machine learning~\cite{BGPS17}.


%

\subsection{The Vadalog System}

Vadalog is a system for performing complex reasoning tasks such as those
required in advanced knowledge graphs~\cite{BeSG18,MFLSS17,FGNS16}. It is Oxford's
contribution to the VADA research project \cite{KKAC17},
a joint effort of the universities of Oxford, Manchester, and Edinburgh,
as well as around 20 industrial partners such as Facebook, BP, and the
NHS (UK national health system).
One of the most fundamental reasoning tasks performed by Vadalog is {\em ontological query answering}: given a database $D$, an ontology $\dep$ (which is essentially a set of logical assertions that allow us to derive new intensional knowledge from $D$), and a query $q(\bar x)$ (typically a conjunctive query), the goal is to compute the certain answers to $q$ w.r.t.~the knowledge base consisting of $D$ and $\dep$, i.e., the tuples of constants $\bar c$ such that, for every relational instance $I \supseteq D$ that satisfies $\dep$, $I$ satisfies the Boolean query $q(\bar c)$ obtained after instantiating $\bar x$ with $\bar c$.
Due to Vadalog's ability to perform ontological query answering, it is currently used as the core deductive database component of the overall Vadalog KGMS, as well as at various industrial partners including the finance, security, and media intelligence industries.

The logical core of the underlying Vadalog language is a rule-based formalism known as {\em warded Datalog$^\exists$}~\cite{GoPi15}, which is a member of the Datalog$^\pm$ family of knowledge representation languages~\cite{CGLMP10}. Warded Datalog$^\exists$ generalizes Datalog with existential quantification in rule heads, and at the same time applies a restriction on how certain ``dangerous'' variables can be used; details are given in Section~\ref{sec:core}. Such a restriction is needed as basic reasoning tasks, e.g., ontological query answering, under arbitrary Datalog$^\exists$ rules become undecidable; see, e.g.,~\cite{BeVa81,CaGK13}.
Let us clarify that Datalog$^\exists$ rules are essentially {\em tuple-generating dependencies} (TGDs) of the form $\forall \bar x \forall \bar y (\varphi(\bar x,\bar y) \ra \exists \bar z \, \psi(\bar x,\bar z))$, where $\varphi$ (the {\em body}) and $\psi$ (the {\em head}) are conjunctions of relational atoms. Therefore, knowledge representation and reasoning should be seen as a modern application of TGDs, which have been introduced decades ago as a unifying framework for database integrity constraints.

The key properties of warded Datalog$^\exists$, which led to its adoption as the logical core on top of which the Vadalog language is built, can be summarized as follows:
\begin{enumerate}
\item {\em Recursion over KGs.} It is able to express full recursion and joins, which are needed to express complex reasoning tasks over KGs. Moreover, navigational capabilities, empowered by recursion, are vital for graph-based structures.

\item {\em Ontological Reasoning over KGs.} After adding a very mild and easy to handle negation, the language is able to express SPARQL reasoning under the OWL 2 QL entailment regime. Recall that SPARQL is the standard language for querying the Semantic Web,\footnote{http://www.w3.org/TR/rdf-sparql-query} while OWL 2 QL is a prominent profile of the OWL 2 Web Ontology Language, the standard formalism for modeling Semantic Web ontologies.\footnote{https://www.w3.org/TR/owl2-overview/}

\item {\em Low Complexity.} Reasoning, in particular, ontological query answering, is tractable (in fact, polynomial time) in data complexity, which is a minimal requirement for allowing scalability over large volumes of data.
\end{enumerate}
Warded Datalog$^\exists$ turned out to be powerful enough for expressing all the tasks given by our industrial partners, while a recent analysis of it focusing on a practical implementation led to the reasoning algorithm around which the Vadalog system is built~\cite{BeSG18}.



\subsection{Research Challenges}

With the aim of isolating more refined formalisms, which will lead to yet more efficient reasoning algorithms, the following fundamental question has emerged in the context of Vadalog:

\medskip


\noindent
\textit{Can we limit the recursion allowed by wardedness in order to obtain a formalism that provides a convenient syntax for expressing useful statements, importantly, most of the scenarios provided by our industrial partners, and at the same time achieves space-efficiency, in particular, {\rm\textsc{NLogSpace}} data complexity?}

\medskip

\noindent Let us stress that \textsc{NLogSpace} data complexity is the best that we can hope for, since navigational capabilities are vital for graph-based structures, and already graph reachability is \textsc{NLogSpace}-hard.
It is known that \textsc{NLogSpace} is contained in the class \textsc{NC}$_2$ of highly parallelizable problems. This means that reasoning in the more refined formalism that we are aiming is principally parallelizable, unlike warded Datalog$^\exists$, which is \textsc{PTime}-complete and intrinsically sequential. Our ultimate goal is to exploit this in the future for the parallel execution of reasoning tasks in both multi-core settings and in the map-reduce model. In fact, we are currently in the process of implementing a multi-core implementation for the refined formalism proposed by the present work.

Extensive benchmark results are available for the Vadalog system, based on a variety of scenarios, both synthetic and industrial scenarios, including: ChaseBench~\cite{BKMM+17}, a benchmark that targets data exchange and query answering problems; iBench, a data exchange benchmark developed at the University of Toronto~\cite{AGCM15}; iWarded, a benchmark specifically targeted at warded sets of TGDs; a DBpedia based benchmark; and a number of other synthetic and industrial scenarios~\cite{BeSG18}.
Let us stress that all the above benchmarks contain only warded sets of TGDs. In fact, a good part of them are not {\em warded by chance}, i.e., they contain joins among ``harmful'' variables, which is one of the distinctive features of wardedness~\cite{BeSG18}.
After analyzing the above benchmarks, we observed that recursion is often used in a restricted way. Approximately 70\% of the TGD-sets use recursion in the following way: the body of a TGD contains at most one atom whose predicate is mutually recursive with a predicate in the head.
More specifically, approximately 55\% of the TGD-sets directly use the above type of recursion, while 15\% can be transformed into warded sets of TGDs that use recursion as explained above. This transformation relies on a standard elimination procedure of unnecessary non-linear recursion.
For example,
\[
\forall x \forall y (E(x,y) \ra T(x,y)) \quad \forall x \forall y \forall z (T(x,y) \wedge T(y,z) \ra  T(x,z)),
\]
which compute the transitive closure of the extensional binary relation $E$ using non-linear recursion, can be rewritten as the set
\[
\forall x \forall y (E(x,y) \ra T(x,y)) \quad \forall x \forall y \forall z (E(x,y) \wedge T(y,z) \ra  T(x,z))
\]
that uses linear recursion.
Interestingly, the type of recursion discussed above has been already studied in the context of Datalog, and is known as {\em piece-wise linear}; see, e.g.,~\cite{AfGT03}. It is a refinement of the well-known {\em linear} recursion~\cite{Naug86,NaSa87}, already mentioned in the above example, which allows only one intensional predicate to appear in the body, while all the other predicates are extensional.

Based on this key observation, the following research questions have immediately emerged:
\begin{enumerate}
\item Does warded Datalog$^\exists$ with piece-wise linear recursion achieve space-efficiency for ontological query answering?\footnote{The idea of combining wardedness with piece-wise linearity has been already mentioned in the invited paper~\cite{BGPS17}, while the obtained formalism is called strongly warded.}

\item Is the combination of wardedness and piece-wise linearity justified? In other words, can we achieve the same with piece-wise linear Datalog$^\exists$ without the wardedness condition?

\item What is the expressiveness of the query language based on warded Datalog$^\exists$ with piece-wise linear recursion relative to prominent query languages such as Datalog?
\end{enumerate}
These are top-priority questions in the context of the Vadalog system since they may provide useful insights towards more efficient reasoning algorithms, in particular, towards parallel execution of reasoning tasks.
The ultimate goal of this work is to analyze piece-wise linearity, and provide definite answers to the above questions.



\subsection{Summary of Contributions}
Our main results can be summarized as follows:

\begin{enumerate}[itemindent=0pt,labelwidth=1em,labelindent=0em,leftmargin=*]
\item Ontological query answering under warded Datalog$^\exists$ with piece-wise linear recursion is \textsc{NLogSpace}-complete in data complexity, and \textsc{PSpace}-complete in combined complexity, which provides a definite answer to our first question. Notice that, as is customary when studying the complexity of ontological query answering, we consider its associated decision problem, where, together with the database $D$, the ontology $\dep$, and the conjunctive query $q$, a tuple $\bar c$ is also part of the input, while the problem is to decided whether $\bar c$ is a certain answer to $q$ w.r.r.~$D$ and $\dep$.
  This is a rather involved result that heavily relies on a novel notion
  of resolution-based proof tree, which is of independent interest. In
  particular, we show that ontological query answering under warded
  Datalog$^\exists$ with piece-wise linear recursion boils down to the
  problem of checking whether a proof tree that enjoys certain
  properties exists, which in turn can be done via a space-bounded
  non-deterministic algorithm. Interestingly, our machinery allows us to
  re-establish the complexity of ontological query answering under
  warded Datalog$^\exists$ via an algorithm that is significantly
  simpler than the one employed in~\cite{GoPi15}. This algorithm is
  essentially the non-determinisitc algorithm for piece-wise linear
  warded Datalog$^\exists$ with the crucial difference that it employs
  alternation.

\item To our surprise, ontological query answering under piece-wise linear Datalog$^\exists$, without the wardedness condition, is undecidable. This result, which is shown via a reduction from the standard unbounded tiling problem, provides a definite answer to our second question: the combination of wardedness and piece-wise linearity is indeed justified.

\item We finally investigate the relative expressive power of the query language based on warded Datalog$^\exists$ with piece-wise linear recursion, which consists of all the queries of the form $Q = (\dep,q)$, where $\dep$ is a warded set of TGDs with piece-wise linear recursion, and $q$ is a conjunctive query, while the evaluation of $Q$ over a database $D$ is precisely the certain answers to $q$ w.r.t.~$D$ and $\dep$. By exploiting our novel notion of proof tree, we show that it is equally expressive to piece-wise linear Datalog. The same approach allows us to elucidate the relative expressiveness of the query language based on warded Datalog$^\exists$ (with arbitrary recursion), showing that it is equally expressive to Datalog.
    We also adopt the more refined notion of program expressive power, introduced in~\cite{ArGP14}, which aims at the decoupling of the set of TGDs and the actual conjunctive query, and show that the query language based on warded Datalog$^\exists$ (with piece-wise linear recursion) is strictly more expressive than Datalog (with piece-wise linear recursion). This result exposes the advantage of value invention that is available in Datalog$^\exists$-based languages.
\end{enumerate}

\noindent
\paragraph{Roadmap.} Preliminaries are given in Section~\ref{sec:preliminaries}. In Section~\ref{sec:core}, we recall the logical core of Vadalog, which in turn relies on the notion of wardedness for TGDs. In Section~\ref{sec:pwl}, we analyze the notion of piece-wise linearity, and show that it achieves space-efficiency for ontological query answering. The formal justification for the combination of wardedness with piece-wise linearity is given in Section~\ref{sec:justification}. In Section~\ref{sec:expressiveness}, we analyze the relative expressiveness of the query languages based on (piece-wise linear) warded sets of TGDs. Finally, in Section~\ref{sec:conclusions}, we give a glimpse on how the current implementation of the Vadalog system is optimized for piece-wise linear
warded sets of TGDs, and describe our future research plans. Selected proofs are deferred to a clearly marked appendix.

\section{Preliminaries}\label{sec:preliminaries}

\paragraph{Basics.} We consider the disjoint countably infinite sets $\ins{C}$, $\ins{N}$, and $\ins{V}$ of {\em constants}, {\em (labeled) nulls}, and {\em variables}, respectively. The elements of $(\ins{C} \cup \ins{N} \cup \ins{V})$ are called {\em terms}. An {\em atom} is an expression of the form $R(\bar t)$, where $R$ is an $n$-ary predicate, and $\bar t$ is an $n$-tuple of terms. We write $\var{\alpha}$ for the set of variables in an atom $\alpha$; this notation extends to sets of atoms. A {\em fact} is an atom that contains only constants.
A {\em substitution} from a set of terms $T$ to a set of terms $T'$ is a function $h \colon T \ra T'$. The restriction of $h$ to a subset $S$ of $T$, denoted $h_{|S}$, is the substitution $\{t \mapsto h(t) \mid t \in S\}$.
A {\em homomorphism} from a set of atoms $A$ to a set of atoms $B$ is a substitution $h$ from the set of terms in $A$ to the set of terms in $B$ such that $h$ is the identity on $\ins{C}$, and $R(t_1,\ldots,t_n) \in A$ implies $h(R(t_1,\ldots,t_n)) =  R(h(t_1),\ldots,h(t_n)) \in B$.
We write $h(A)$ for the set of atoms $\{h(\alpha) \mid \alpha \in A\}$.
For brevity, we may write $[n]$ for the set $\{1,\ldots,n\}$, where $n \geq 0$.

\medskip
\noindent
\paragraph{Relational Databases.} A {\em schema} $\ins{S}$ is a finite
set of relation symbols (or predicates), each having an associated
\emph{arity}. We write $R/n$ to denote that $R$ has arity $n \geq 0$. A
position $R[i]$ in $\ins{S}$, where $R/n \in \ins{S}$ and $i \in [n]$,
identifies the $i$-th argument of $R$. An {\em instance} over $\ins{S}$
is a (possibly infinite) set of atoms over $\ins{S}$ that contain
constants and nulls, while a {\em database} over $\ins{S}$ is a finite
set of facts over $\ins{S}$. The {\em active domain} of an instance $I$,
denoted $\adom{I}$, is the set of all terms occurring in $I$.

\medskip

\noindent
\paragraph{Conjunctive Queries.}
A {\em conjunctive query} (CQ) over $\ins{S}$ is a first-order formula of the form
\[
q(\bar x) \ \coloneqq
\
\exists \bar y\, \big(R_1(\bar z_1) \wedge \dots \wedge R_n(\bar
z_n)\big),
\]
where each $R_i(\bar z_i)$, for $ i \in [n]$, is an atom without nulls
over $\ins{S}$, each variable mentioned in the $\bar z_i$'s
appears either in $\bar x$ or $\bar y$, and $\bar x$ are the {\em output variables} of $q$.
%
For convenience, we adopt the rule-based syntax of CQs, i.e., a CQ as the one above will be written as the rule
\[
Q(\bar x)\ \la\ R_1(\bar z_1), \dots, R_n(\bar z_n),
\]
where $Q$ is a predicate used only in the head of CQs. We write $\atoms{q}$ for the set of atoms $\{R_1(\bar z_1), \dots, R_n(\bar z_n)\}$.
The {\em evaluation} of $q(\bar x)$ over an instance $I$, denoted $q(I)$, is
the set of all tuples $h(\bar x)$ of constants with $h$ being a homomorphism from $\atoms{q}$ to $I$.
%

\medskip

\noindent
\paragraph{Tuple-Generating Dependencies.} A {\em tuple-generating dependency} (TGD) $\sigma$ is a first-order sentence
\[
\forall \bar x \forall \bar y \left(\phi(\bar x,\bar y) \ra \exists \bar z\, \psi(\bar x,\bar z)\right),
\]
where $\bar x, \bar y, \bar z$ are tuples of variables of $\ins{V}$, and $\phi,\psi$ are conjunctions of atoms without constants and nulls.
For brevity, we write $\sigma$ as $\phi(\bar x,\bar y) \ra \exists \bar z\, \psi(\bar x,\bar z)$, and use comma instead of $\wedge$ for joining atoms. We refer to $\phi$ and $\psi$ as the {\em body} and {\em head} of $\sigma$, denoted $\body{\sigma}$ and $\head{\sigma}$, respectively.
The {\em frontier} of the TGD $\sigma$, denoted $\fr{\sigma}$, is the set of variables that appear both in the body and the head of $\sigma$.
We also write $\mathsf{var}_{\exists}(\sigma)$ for the existentially quantified variables of $\sigma$.
The schema of a set $\dep$ of TGDs, denoted $\sch{\dep}$, is the set of predicates occurring in $\dep$.
An instance $I$ satisfies a TGD $\sigma$ as the one above, written $I \models \sigma$, if the following holds: whenever there exists a homomorphism $h$ such that $h(\phi(\bar x, \bar y)) \subseteq I$, then there exists $h' \supseteq h_{|\bar x}$ such that $h'(\psi(\bar x,\bar z)) \subseteq I$.\footnote{By abuse of notation, we sometimes treat a tuple of variables as a set of variables, and a conjunction of atoms as a set of atoms.}
The instance $I$ satisfies a set $\dep$ of TGDs, written $I \models \dep$, if $I \models \sigma$ for each $\sigma \in \dep$.
%

\medskip
\noindent
\paragraph{Query Answering under TGDs.} The main reasoning task under TGD-based languages is \emph{conjunctive query answering}. Given a database $D$ and a set $\dep$ of TGDs, a {\em model} of $D$ and $\dep$ is an instance $I$ such that $I \supseteq D$ and $I \models \dep$.
Let $\mods{D}{\dep}$ be the set of all models of $D$ and $\dep$.
The {\em certain answers} to a CQ $q$ w.r.t.~$D$ and $\dep$ is
\[
\cert{q}{D}{\dep}\ \coloneqq\ \bigcap \set{q(I) \mid I \in \mods{D}{\dep}}.
\]
Our main task is to compute the certain answers to a CQ w.r.t.~a database and a set of TGDs from a certain class $\class{C}$ of TGDs; concrete classes of TGDs are discussed below. As is customary when studying the complexity of this problem, we focus on its decision version:

\medskip

\begin{center}
\fbox{\begin{tabular}{ll}
{\small PROBLEM} : & $\qans(\class{C})$
\\
{\small INPUT} : & A database $D$, a set $\dep \in \class{C}$ of TGDs,\\
& a CQ $q(\bar x)$, and a tuple ${\bar c} \in \adom{D}^{|{\bar x}|}$.
\\
{\small QUESTION} : &  Is it the case that ${\bar c} \in \cert{q}{D}{\dep}$?
\end{tabular}}
\end{center}

\medskip

\noindent We consider the standard complexity measures: {\em combined
  complexity} and {\em data complexity}, where the latter measures the
complexity of the problem assuming that the set of TGDs and the
CQ are fixed.

A useful algorithmic tool for tackling the above problem is the well-known  {\em chase procedure}; see, e.g.,~\cite{CaGK13,FKMP05,JoKl84,MaMS79}. We start by defining a single chase step. Let $I$ be an instance and $\sigma = \phi(\bar x,\bar y)
\rightarrow \exists \bar z \, \psi(\bar x,\bar z)$ a TGD.
We say that $\sigma$ is \emph{applicable} w.r.t.~$I$ if there exists a homomorphism $h$ such that $h(\phi(\bar x,\bar y)) \subseteq I$. In this case, {\em the result of applying $\sigma$ over $I$ with $h$} is the instance $J = I \cup \{h'(\psi(\bar x,\bar z))\}$, where $h'(z)$ is a fresh null not occurring in $I$, for every $z \in \bar z$. Such a single chase step is denoted $I \app{\sigma}{h} J$.
Consider now an instance $I$, and a set $\dep$ of TGDs. A {\em chase sequence for $I$ under $\dep$} is a sequence $(I_i \app{\sigma_i}{h_i} I_{i+1})_{i \geq 0}$
of chase steps such that: (1) $I = I_0$; (2) for each $i \geq 0$, $\sigma_i \in \dep$; and (3) $\bigcup_{i \geq 0} I_i \models \dep$. We call $\bigcup_{i \geq 0} I_i$ the {\em result} of this chase sequence, which always exists. Although the result of a chase sequence is not necessarily unique (up to isomorphism), each such result is equally useful for query answering purposes, since it can be homomorphically embedded into every other result. Hence, we denote by $\chase{I}{\dep}$ the result of an arbitrary chase sequence for $I$ under $\dep$.
The following is a classical result:

\begin{proposition}\label{pro:chase}
Given a database $D$, a set $\dep$ of TGDs, and a CQ $q$, $\cert{q}{D}{\dep} = q(\chase{D}{\dep})$.
\end{proposition}

\section{The Logical Core of VADALOG}\label{sec:core}
%


A crucial component of the Vadalog system is its reasoning engine, which in turn is built around the Vadalog language, a general-purpose formalism for knowledge representation and reasoning. The logical core of this language is the well-behaved class of warded sets of TGDs that has been proposed in~\cite{GoPi15}.

\medskip
\noindent
\paragraph{An Intuitive Description.}
Wardedness applies a syntactic restriction on how certain ``dangerous''
variables of a set of TGDs are used. These are body variables that can
be unified with a null during the chase, and that are also propagated to
the head.
For example, given
\[
P(x) \ra \exists z \, R(x,z) \quad \text{ and } \quad R(x,y) \ra P(y)
\]
the variable $y$ in the body of the second TGD is dangerous. Indeed, once the chase applies the first TGD, an atom of the form $R(\_,\bot)$ is generated, where $\bot$ is a null value, and then the second TGD is triggered with the variable $y$ being unified with $\bot$ that is propagated to the obtained atom $P(\bot)$.
It has been observed that the liberal use of dangerous variables leads to a prohibitively high computational complexity of the main reasoning tasks, in particular of CQ answering~\cite{CaGK13}.
The main goal of wardedness is to limit the use of dangerous variables with the aim of taming the way that null values are propagated during the execution of the chase procedure. This is achieved by posing the following conditions:
\begin{enumerate}
\item all the dangerous variables should appear together in a single body atom $\alpha$, called a ward, and

\item $\alpha$ can share only harmless variables with the rest of the body, i.e., variables that unify only with constants.
\end{enumerate}
We proceed to formalize the above description.

\medskip
\noindent
\paragraph{The Formal Definition.} We first need some auxiliary notions. The set of positions of a schema $\ins{S}$, denoted $\pos{\ins{S}}$, is defined as $\{R[i] \mid R/n \in \ins{S}, \text{ with } n \geq 1, \text{ and } i \in [n]\}$. Given a set $\dep$ of TGDs, we write $\pos{\dep}$ instead of $\pos{\sch{\dep}}$. The set of {\em affected positions} of $\sch{\dep}$, denoted $\aff{\dep}$, is inductively defined as follows:
\begin{itemize}
\item[--] if there exists $\sigma \in \dep$ and a variable $x \in \mathsf{var}_{\exists}(\sigma)$ at position $\pi$, then $\pi \in \aff{\dep}$, and

\item[--] if there exists $\sigma \in \dep$ and a variable $x \in \fr{\sigma}$ in the body of $\sigma$ only at positions of $\aff{\dep}$, and $x$ appears in the head of $\sigma$ at position $\pi$, then $\pi \in \aff{\dep}$.
\end{itemize}
Let $\nonaff{\dep} = \pos{\dep} \setminus \aff{\dep}$.
We can now classify the variables in the body of a TGD into
{harmless}, {harmful}, and {dangerous}. Fix a TGD
$\sigma \in \dep$ and a variable $x$ in $\body{\sigma}$:
\begin{itemize}
\item[--] $x$ is {\em harmless} if at least one occurrence of it appears in $\body{\sigma}$ at a position of $\nonaff{\dep}$,

\item[--] $x$ is {\em harmful} if it is not harmless, and

\item[--] $x$ is {\em dangerous} if it is harmful and belongs to $\fr{\sigma}$.
\end{itemize}

We are now ready to formally introduce wardedness.

\begin{definition}[Wardedness]
  \label{def:warded-tgds}
A set $\dep$ of TGDs is {\em warded} if, for each TGD $\sigma \in \dep$, there are no dangerous variables in $\body{\sigma}$, or there exists an atom $\alpha \in \body{\sigma}$, called a \emph{ward}, such that:
\begin{enumerate}
\item[--] all the dangerous variables in $\body{\sigma}$ occur in $\alpha$, and
\item[--] each variable of $\var{\alpha} \cap \var{\body{\sigma} \setminus \{\alpha\}}$ is harmless.
\end{enumerate}
We denote by $\class{WARD}$ the class of all (finite) warded sets of
TGDs. \hfill\markfull
\end{definition}

The problem of CQ answering under warded sets of TGDs has been recently investigated in~\cite{GoPi15}:

\begin{proposition}\label{the:warded-complexity}
$\qans(\class{WARD})$ is {\rm \textsc{ExpTime}}-complete in combined complexity, and {\rm \textsc{PTime}}-complete in data complexity.
\end{proposition}

Note that~\cite{GoPi15} deals only with data complexity. However, it is implicit that the same algorithm provides an \textsc{ExpTime} upper bound in combined complexity, while the lower bounds are inherited from Datalog since a set of Datalog rules (seen as TGDs) is warded.

\medskip
\noindent
\paragraph{A Key Application.} One of the distinctive features of wardedness, which is crucial for the purposes of the Vadalog system, is the fact that it can express every SPARQL query under the OWL 2 QL direct semantics entailment regime, which is inherited from the OWL 2 direct semantics entailment regime; for details, see~\cite{ArGP14,GoPi15}.
Recall that SPARQL is the standard language for querying the Semantic Web,\footnote{http://www.w3.org/TR/rdf-sparql-query} while OWL 2 QL is a prominent profile of the OWL 2 Web Ontology Language, the standard formalism for modeling Semantic Web ontologies.\footnote{https://www.w3.org/TR/owl2-overview/}
We give a simple example of a warded set of TGDs, which is extracted from the set of TGDs that encodes the OWL 2 direct semantics entailment regime for OWL 2 QL.

\begin{example}\label{exa:warded-tgds}
An OWL 2 QL ontology can be stored in a database using atoms of the form ${\rm Restriction}(c,p)$ stating that the class $c$ is a restriction of the property $p$, ${\rm SubClass}(c,c')$ stating that $c$ is a subclass of $c'$, and ${\rm Inverse}(p,p')$ stating that $p$ is the inverse property of $p'$. We can then compute all the logical inferences of the given ontology using TGDs as the ones below:
\begin{align*}
{\rm SubClass}(x,y)\ &\ra\ {\rm SubClass}^\star(x,y)\\
{\rm SubClass}^\star(x,y),{\rm SubClass}(y,z)\ &\ra\ {\rm SubClass}^\star(x,z)\\
\underline{{{\rm Type}(x,y)}}, {\rm SubClass}^\star(y,z)\ &\ra\ {\rm Type}(x,z)\\
\underline{{{\rm Type}(x,y)}}, {\rm Restriction}(y,z)\ &\ra\ \exists w \, {\rm Triple}(x,z,w)\\
\underline{{{\rm Triple}(x,y,z)}}, {\rm Inverse}(y,w)\ &\ra\ {\rm Triple}(z,w,x)\\
\underline{{{\rm Triple}(x,y,z)}}, {\rm Restriction}(w,y)\ &\ra\ {\rm Type}(x,w).
\end{align*}
The first two TGDs are responsible for computing the transitive closure of the ${\rm SubClass}$ relation, while the third TGD transfers the class type, i.e., if $a$ is of type $b$ and $b$ is a subclass of $c$, then $a$ is also of type $c$. Moreover, the fourth TGD states that if $a$ is of type $b$ and $b$ is the restriction of the property $p$, then $a$ is related to some $c$ via the property $p$, which is encoded by the atom ${\rm Triple}(a,p,c)$. Analogously, the last two TGDs encode the usual meaning of inverses and the effect of restrictions on types.

It is easy to verify that the above set of TGDs is warded, where the underlined atoms are the wards; if no atom is underlined, then there are no dangerous variables. A variable in an atom with predicate ${\rm Restriction}$, ${\rm SubClass}$, ${\rm SubClass}^\star$, or ${\rm Inverse}$, is trivially harmless. The frontier variables that appear at ${\rm Type}[1]$, ${\rm Triple}[1]$, or ${\rm Triple}[3]$, are dangerous, and the underlined atoms are acting as wards. \hfill\markfull
\end{example}


\section{Limiting Recursion}
\label{sec:pwl}
\label{SEC:PWL}

We now focus on our main research question: can we limit the recursion allowed by wardedness in order to obtain a formalism that provides a convenient syntax for expressing useful recursive statements, and at the same time achieve space-efficiency?
The above question has been extensively studied in the 1980s for Datalog programs, with {\em linear Datalog} being a key fragment that achieves a good balance between expressivity and complexity; see, e.g.,~\cite{Naug86,NaSa87}. A Datalog program $\dep$ is \emph{linear} if, for each rule in $\dep$, its body contains at most one intensional predicate, i.e., a predicate that appears in the head of at least one rule of $\dep$. In other words, linear Datalog allows only for linear recursion, which is able to express many real-life recursive queries. However, for our purposes, linear recursion does not provide the convenient syntax that we are aiming at. Already the simple set of TGDs in Example~\ref{exa:warded-tgds}, which is part of the set of TGDs that encodes the OWL 2 direct semantics entailment regime for OWL 2 QL, uses non-linear recursion due to
\[
{\rm Type}(x,y), {\rm SubClass}^\star(y,z) \ra {\rm Type}(x,z),
\]
where both body atoms have intensional predicates.
Moreover, after analyzing several real-life examples of warded sets of TGDs, provided by our industrial partners, we observed that the employed recursion goes beyond linear recursion.
On the other hand, the set of TGDs in Example~\ref{exa:warded-tgds}, as well as most of the examples came from our industrial partners, use recursion in a restrictive way: each TGD has at most one body atom whose predicate is mutually recursive with a predicate occurring in the head of the TGD.
Interestingly, this more liberal version of linear recursion has been already investigated in the context of Datalog, and it is known as {\em piece-wise linear}; see, e.g.,~\cite{AfGT03}.
Does this type of recursion lead to the space-efficient fragment of warded sets of TGDs that we are looking for? The rest of this section is devoted to showing this rather involved result.


Let us start by formally defining the class of piece-wise linear sets of TGDs. To this end, we need to define when two predicates are mutually recursive, which in turn relies on the well-known notion of the predicate graph.
The {\em predicate graph} of a set $\dep$ of TGDs, denoted $\pg{\dep}$, is a directed graph $(V,E)$, where $V = \sch{\dep}$, and there exists an edge from a predicate $P$ to a predicate $R$, i.e., $(P,R) \in E$, iff there exists a TGD $\sigma \in \dep$ such that $P$ occurs in $\body{\sigma}$ and $R$ occurs in $\head{\sigma}$.
Two predicates $P,R \in \sch{\dep}$ are {\em mutually recursive} (w.r.t.~$\dep$) if there exists a cycle in $\pg{\dep}$ that contains both $P$ and $R$ (i.e., $R$ is reachable from $P$, and vice versa). We are now ready to define piece-wise linearity for TGDs.

\begin{definition}[Piece-wise Linearity]\label{def:pwl-tgds}
A set $\dep$ of TGDs is {\em piece-wise linear} if, for each TGD $\sigma \in \dep$, there exists at most one atom in $\body{\sigma}$ whose predicate is mutually recursive with a predicate in $\head{\sigma}$. Let $\class{PWL}$ be the class of piece-wise linear sets of TGDs. \hfill\markfull
\end{definition}

The main result of this section follows:

\begin{theorem}\label{the:warded-pwl-complexity}
$\qans(\class{WARD} \cap \class{PWL})$ is {\rm \textsc{PSpace}}-complete in combined complexity, and {\rm \textsc{NLogSpace}}-complete in data complexity.
\end{theorem}

The lower bounds are inherited from linear Datalog. The difficult task is to establish the upper bounds. This relies on a novel notion of proof tree, which is of independent interest.
As we shall see, our notion of proof tree leads to space-bounded algorithms that allow us to show the upper bounds in Theorem~\ref{the:warded-pwl-complexity}, and also re-establish in a transparent way the upper bounds in Proposition~\ref{the:warded-complexity}.
Moreover, in Section~\ref{sec:expressiveness}, we are going to use proof trees for studying the relative expressive power of (piece-wise linear) warded sets of TGDs.

\subsection{Query Answering via Proof Trees}

It is known that given a CQ $q$ and a set $\dep$ of TGDs, we can unfold $q$ using the TGDs of $\dep$ into an infinite union of CQs $q_\dep$ such that, for every database $D$, $\cert{q}{D}{\dep} = q_\dep(D)$; see, e.g.,~\cite{GoOP14,KoLMT15}. Let us clarify that in our context, an unfolding, which is essentially a resolution step, is more complex than in the context of Datalog due to the existentially quantified variables in the head of TGDs.
The intention underlying our notion of proof tree is to encode in a tree the sequence of CQs, generated during the unfolding of $q$ with $\dep$, that leads to a certain CQ $q'$ of $q_\dep$ in such a way that each intermediate CQ, as well as $q'$, is carefully decomposed into smaller subqueries that form the nodes of the tree, while the root corresponds to $q$ and the leaves to $q'$.
As we shall see, if we focus on well-behaved classes of TGDs such as (piece-wise linear) warded sets of TGDs, we can establish upper bounds on the size of these subqueries, which in turn allow us to devise space-bounded algorithms for query answering.
In what follows, we define the notion of proof tree (Definition~\ref{def:proof-tree}), and establish its correspondence with query answering (Theorem~\ref{the:proof-trees-qans}). To this end, we need to introduce the main building blocks of a proof tree: chunk-based resolution (Definition~\ref{def:resolution}), a query decomposition technique (Definition~\ref{def:decomposition}), and the notion of specialization for CQs (Definition~\ref{def:specialization}).

\medskip

\noindent
\paragraph{Chunk-based Resolution.} Let $A$ and $B$ be non-empty set of atoms that mention only constants and variables. The sets $A$ and $B$ \emph{unify} if there is a substitution $\gamma$, which is the identity on $\ins{C}$, called \emph{unifier for $A$ and $B$}, such that $\gamma(A) = \gamma(B)$. A \emph{most general unifier} (MGU) for $A$ and $B$ is a unifier for $A$ and $B$, denoted $\gamma_{A,B}$, such that, for each unifier $\gamma$ for $A$ and $B$, $\gamma = \gamma' \circ \gamma_{A,B}$ for some substitution $\gamma'$. Notice that if two sets of atoms unify, then there exists always a MGU, which is unique (modulo variable renaming).

Given a CQ $q(\bar x)$ and a set of atoms $S \subseteq \atoms{q}$, we call a variable $y \in \var{S}$ {\em shared} if $y \in \bar x$, or $y \in \var{\atoms{q} \setminus S}$.
A {\em chunk unifier} of $q$ with a TGD $\sigma$ (where $q$ and $\sigma$ do not share variables) is a triple $(S_1,S_2,\gamma)$, where $\emptyset \subset S_1 \subseteq \atoms{q}$, $\emptyset \subset S_2 \subseteq \head{\sigma}$, and $\gamma$ is a unifier for $S_1$ and $S_2$ such that, for each $x \in \var{S_2} \cap \mathsf{var}_{\exists}(\sigma)$,
\begin{enumerate}
\item $\gamma(x) \not\in \ins{C}$, i.e., $\gamma(x)$ is not constant, and
\item $\gamma(x) = \gamma(y)$ implies $y$ occurs in $S_1$ and is not shared.
\end{enumerate}
The chunk unifier $(S_1,S_2,\gamma)$ is {\em most general} (MGCU) if $\gamma$ is an MGU for $S_1$ and $S_2$.
Notice that the the variables of $\mathsf{var}_{\exists}(\sigma)$ occurring in $S_2$ unify (via $\gamma$) only with non-shared variables of $S_1$. This ensures that $S_1$ is a ``chunk'' of $q$ that can be resolved as a whole via $\sigma$ using $\gamma$.\footnote{A similar notion known as piece unifier has been defined in~\cite{KoLMT15}.}
Without the additional conditions on the substitution $\gamma$, we may get unsound resolution steps. Consider, e.g., the CQ and TGD
\[
Q(x) \la R(x,y),S(y) \quad \text{ and } \quad  P(x') \ra \exists y' \,  R(x',y').
\]
Resolving the atom $R(x,y)$ in the query with the given TGD using the substitution $\gamma = \{x \mapsto x',y \mapsto y'\}$  would be an unsound step since the shared variable $y$ is lost. This is because $y'$ is unified with the shared variable $y$. On the other hand, $R(x,y),S(y)$ can be resolved with the TGD $\sigma = P(x') \ra \exists y' \, R(x',y'),S(y')$ using $\gamma$; in fact, the chunk unifier is $(\atoms{q},\head{\sigma},\gamma)$.
%

\begin{definition}[Chunk-based Resolution]
\label{def:resolution}
Let $q(\bar x)$ be a CQ and $\sigma$ a TGD. A {\em $\sigma$-resolvent} of $q$ is a CQ $q'(\gamma(\bar x))$ with $\body{q'} = \gamma((\atoms{q} \setminus S_1) \cup \body{\sigma})$ for a MGCU $(S_1,S_2,\gamma)$ of $q$ with $\sigma$. \hfill\markfull
\end{definition}





\noindent
\paragraph{Query Decomposition.}
As discussed above, the purpose of a proof tree is to encode a finite branch of the unfolding of a CQ $q$ with a set $\dep$ of TGDs. Such a branch is a sequence $q_0,\ldots,q_n$ of CQs, where $q = q_0$, while, for each $i \in [n]$, $q_i$ is a $\sigma$-resolvent of $q_{i-1}$ for some $\sigma \in \dep$.
One may think that the proof tree that encodes the above branch is the finite labeled path $v_0,\ldots,v_n$, where each $v_i$ is labeled by $q_i$.
However, another crucial goal of such a proof tree, which is not achieved via the naive path encoding, is to split each resolvent $q_i$, for $i > 0$, into smaller subqueries $q_{i}^{1},\ldots,q_{i}^{n_i}$, which are essentially the children of $q_i$, in such a way that they can be processed independently by resolution.
The crux of this encoding is that it provides us with a mechanism for keeping the CQs that must be processed by resolution small.


The key question here is how a CQ $q$ can be decomposed into subqueries that can be processed independently. The subtlety is that, after splitting $q$, occurrences of the same variable may be separated into different subqueries. Thus, we need a way to ensure that a variable in $q$, which appears in different subqueries after the splitting, is indeed treated as the same variable, i.e., it has the same meaning.
We deal with this issue by restricting the set of variables in $q$ of which occurrences can be separated during the splitting step. In particular, we can only separate occurrences of an output variable.
This relies on the convention that output variables correspond to fixed constant values of $\ins{C}$, and thus their name is ``freezed'' and never renamed by subsequent resolution steps. Hence, we can separate occurrences of an output variable into different subqueries, i.e., different branches of the proof tree, without losing the connection between them.

Summing up, the idea underlying query decomposition is to split the CQ at hand into smaller subqueries that keep together all the occurrences of a non-output variable, but with the freedom of separating occurrences of an output variable.

\begin{definition}[Query Decomposition]
\label{def:decomposition}
Given a CQ $q(\bar x)$, a {\em decomposition} of $q$ is a set of CQs $\{q_1(\bar y_1),\ldots,q_n(\bar y_n)\}$, where $n \geq 1$ and $\bigcup_{i \in [n]} \atoms{q_i} = \atoms{q}$, such that, for each $i \in [n]$:
\begin{enumerate}
\item $\bar y_i$ is the restriction of $\bar x$ on the variables in $q_i$, and

\item for every $\alpha,\beta \in \atoms{q}$, if $\alpha \in \atoms{q_i}$ and $\var{\alpha} \cap \var{\beta} \not\subseteq \bar x$, then $\beta \in \atoms{q_i}$. \hfill\markfull
\end{enumerate}
\end{definition}

%

\noindent
\paragraph{Query Specialization.} From the above discussion, one expects that a proof tree of a CQ $q$ w.r.t.~a set $\dep$ of TGDs can be constructed by starting from $q$, which is the root, and applying two steps: resolution and decomposition.
Unfortunately, this is not enough for our purposes as we may run into the following two problems:
(i) we may lose vital resolution steps because two output variables may correspond to the same constant value, and thus a unifier will be forced to unify them, but this is forbidden due to the convention discussed above, i.e., output variables keep their names, and
(ii) some of the subqueries will mistakenly remain large since we have no way to realize that a non-output variable corresponds to a fixed constant value, which in turn allows us to ``freeze'' its name and separate different occurrences of it during the decomposition step.

The above issues can be solved by having an intermediate step between resolution and decomposition, the so-called specialization step.
A specialization of a CQ is obtained by converting some non-output variables of it into output variables, while keeping their name, or taking the name of an output variable.

\begin{definition}[Query Specialization]
  \label{def:specialization}
Let $q(\bar x)$ be a CQ with $\atoms{q} = \{\alpha_1,\ldots,\alpha_n\}$. A {\em specialization} of $q$ is a CQ
\[
Q(\bar x,\bar y)\ \la \rho_{\bar z}(\alpha_1,\ldots,\alpha_n)
\]
where $\bar y,\bar z$ are (possibly empty) disjoint tuples of non-output variables of $q$, and $\rho_{\bar z}$ is a substitution from $\bar z$ to $\bar x \cup \bar y$. \hfill\markfull
\end{definition}

\noindent
\paragraph{Proof Trees.}
We are now ready to introduce our new notion of proof tree. But let us first fix some notation. Given a partition $\pi = \{S_1,\ldots,S_m\}$ of a set of variables, we write $\eq_\pi$ for the substitution that maps the variables of $S_i$ to the same variable $x_i$, where $x_i$ is a distinguished element of $S_i$.
We should not forget the convention that output variables cannot be renamed, and thus, a resolution step should use a MGCU that preserves the output variables. In particular, given a CQ $q$ and a TGD $\sigma$, a $\sigma$-resolvent of $q$ is called {\em IDO} if the underlying MGCU uses a substitution that is the identity on the output variables of $q$ (hence the name IDO).
Finally, given a TGD $\sigma$ and some arbitrary object $o$ (e.g., $o$ can be the node of a tree, or an integer number), we write $\sigma_o$ for the TGD obtained by renaming each variable $x$ in $\sigma$ into $x_o$. This is a simple mechanism for uniformly renaming the variables of a TGD in order to avoid undesirable clatter among variables during a resolution step.

\begin{definition}[Proof Tree]
\label{def:proof-tree}
Let $q(\bar x)$ be a CQ with $\atoms{q} = \{\alpha_1,\ldots,\alpha_n\}$, and $\dep$ a set of TGDs.
A {\em proof tree} of $q$ w.r.t.~$\dep$ is a triple $\mathcal{P} = (T,\lambda,\pi)$, where $T = (V,E)$ is a finite rooted tree, $\lambda$ a labeling function that assigns a CQ to each node of $T$, and $\pi$ a partition of $\bar x$, such that, for each node $v \in V$:
\begin{enumerate}
\item If $v$ is the root node of $T$, then $\lambda(v)$ is the CQ $Q(\eq_\pi(\bar x)) \la \eq_\pi(\alpha_1,\ldots,\alpha_m)$.

\item If $v$ has only one child $u$, $\lambda(u)$ is an IDO $\sigma_v$-resolvent of $\lambda(v)$ for some $\sigma \in \dep$, or a specialization of $\lambda(v)$.

\item If $v$ has $k > 1$ children $u_1,\ldots,u_k$, then $\{\lambda(u_1),\ldots,\lambda(u_k)\}$ is a decomposition of $\lambda(v)$.
\end{enumerate}
Assuming that $v_1,\ldots,v_m$ are the leaf nodes of $T$, the {\em CQ induced by} $\mathcal{P}$ is defined as
\[
Q(\eq_\pi(\bar x))\ \la\ \alpha_1,\ldots,\alpha_\ell,
\]
where $\{\alpha_1,\ldots,\alpha_\ell\} = \bigcup_{i \in [m]} \atoms{\lambda(v_i)}$. \hfill\markfull
\end{definition}

%

The purpose of the partition $\pi$ is to indicate that some output variables correspond to the same constant value -- this is why variables in the same set of $\pi$ are unified via the substitution $\eq_\pi$. This unification step is crucial in order to safely use, in subsequent resolution steps, substitutions that are the identity on the output variables. If we omit this initial unification step, we may lose important resolution steps, and thus being incomplete for query answering purposes.
The main result of this section, which exposes the connection between proof trees and CQ answering, follows. By abuse of notation, we write $\ca{P}$ for the CQ induced by $\mathcal{P}$.

\begin{theorem}\label{the:proof-trees-qans}
Consider a database $D$, a set $\dep$ of TGDs, a CQ $q(\bar x)$, and $\bar c \in \adom{D}^{|\bar x|}$. The following are equivalent:
\begin{enumerate}
\item $\bar c \in \cert{q}{D}{\dep}$.
\item There exists a proof tree $\mathcal{P}$ of $q$ w.r.t.~$\dep$ such that $\bar c \in \mathcal{P}(D)$.
\end{enumerate}
\end{theorem}

The proof of the above result relies on the soundness and completeness of chunk-based resolution. Given a set $\dep$ of TGDs and a CQ $q(\bar x)$, by exhaustively applying chunk-based resolution, we can construct a (possibly infinite) union of CQs $q_\dep$ such that, for every database $D$, $\cert{q}{D}{\dep} = q_\dep(D)$; implicit in~\cite{GoOP14,KoLMT15}. In other words, given a tuple $\bar c \in \adom{D}^{|\bar x|}$, $\bar c \in \cert{q}{D}{\dep}$ iff there exists a CQ $q'(\bar x)$ in $q_\dep$ such that $\bar c \in q'(D)$. It is now not difficult to show that the later statement is equivalent to the existence of a proof tree $\mathcal{P}$ of $q$ w.r.t.~$\dep$ such that $\bar c \in \mathcal{P}(D)$, and the claim follows.

\subsection{Well-behaved Proof Trees}

Theorem~\ref{the:proof-trees-qans} states that checking whether a tuple $\bar c$ is a certain answer boils down to deciding whether there exists a proof tree $\cali{P}$ such that $\bar c$ is an answer to the CQ induced by $\cali{P}$ over the given database. Of course, the latter is an undecidable problem in general. However, if we focus on (piece-wise linear) warded sets of TGDs, it suffices to check for the existence of a well-behaved proof tree with certain syntactic properties, which in turn allows us to devise a decision procedure.
We proceed to make this more precise.
For technical clarity, we assume, w.l.o.g., TGDs with only one atom in the head since we can always convert a warded set of TGDs into one with single-atom heads, while certain answers are preserved; for the transformation see, e.g.,~\cite{CaGP12}.

\medskip

\noindent
\paragraph{Piece-wise Linear Warded Sets of TGDs.}
For piece-wise linear warded sets of TGDs, we can strengthen Theorem~\ref{the:proof-trees-qans} by focussing on a certain class of proof trees that enjoy two syntactic properties: (i) they have a path-like structure, and (ii) the size of the CQs that label their nodes is bounded by a polynomial.
The first property is formalized via linear proof trees. Let $\mathcal{P} = (T,\lambda,\pi)$, where $T = (V,E)$, be a proof tree of a CQ $q$ w.r.t.~a set $\dep$ of TGDs. We call $\mathcal{P}$ {\em linear} if, for each node $v \in V$, there exists at most one node $u \in V$ such that $(v,u) \in E$ and $u$ is not a leaf in $T$, i.e., $v$ has at most one child that is not a leaf.
The second property relies on the notion of node-width of a proof
tree. Formally, the {\em node-width} of $\mathcal{P}$ is
\[
\nwd{\mathcal{P}}\ \coloneqq\ \max_{v \in V}  \{|\lambda(v)|\},
\]
i.e., the size of the largest CQ that labels a node of $T$.

Before we strengthen Theorem~\ref{the:proof-trees-qans}, let us define the polynomial that will allow us to bound the node-width of the linear proof trees that we need to consider. This polynomial relies on the notion of predicate level. Consider a set $\dep$ of TGDs. For a predicate $P \in \sch{\Sigma}$, we write $\rec{P}$ for the set of predicates of $\sch{\dep}$ that are mutually recursive to $P$ according to $\pg{\dep} = (V,E)$.
Let $\ell_\Sigma \colon \sch{\Sigma} \ra \mathbb{N}$ be the unique function that satisfies
\begin{align*}
  \lvli{P}{\Sigma}\ =\ \max\set{\lvli{R}{\Sigma} \mid \tup{R,P} \in E, R \not\in \rec{P}} + 1,
\end{align*}
with $\lvli{P}{\Sigma}$ being the {\em level} (w.r.t.~$\dep$) of $P$, for each $P \in \sch{\Sigma}$.
%
We can now define the polynomial
\[
f_{\class{WARD} \cap \class{PWL}}(q,\dep) \coloneqq (|q| + 1) \cdot \max_{\mathclap{P \in \sch{\dep}}}\ \set{\lvli{P}{\Sigma}} \cdot \max_{\sigma \in \dep} \{|\body{\sigma}|\}.
\]

We can now strengthen Theorem~\ref{the:proof-trees-qans}. But let us first clarify that, in the case of piece-wise linear warded sets of TGDs, apart from only one atom in the head, we also assume, w.l.o.g., that the level of a predicate in the body of TGD $\sigma$ is $k$ or $k-1$, where $k$ is the level of the predicate in the head of $\sigma$. The following holds:

\begin{theorem}\label{the:proof-trees-qans-pwl-warded}
Consider a database $D$, a set $\dep \in \class{WARD} \cap \class{PWL}$ of TGDs, a CQ $q(\bar x)$, and $\bar c \in \adom{D}^{|\bar x|}$. The following are equivalent:
\begin{enumerate}
\item $\bar c \in \cert{q}{D}{\dep}$.
\item There is a linear proof tree $\mathcal{P}$ of $q$ w.r.t.~$\dep$ with $\nwd{\cali{P}} \leq f_{\class{WARD} \cap \class{PWL}}(q,\dep)$ such that $\bar c \in \mathcal{P}(D)$.
\end{enumerate}
\end{theorem}



\noindent
\paragraph{Warded sets of TGDs.}
Now, in the case of arbitrary warded sets of TGDs, we cannot focus only
on linear proof trees. Nevertheless, we can still bound the node-width
of the proof trees that we need to consider by the following polynomial,
which, unsurprisingly, does not rely anymore on the notion of predicate
level:
\[
f_{\class{WARD}}(q,\dep) \coloneqq 2 \cdot \max\left\{|q|,\max_{\sigma \in \dep}  \{|\body{\sigma}|\}\right\}.
\]
Theorem~\ref{the:proof-trees-qans} can be strengthened as follows:

%

\begin{theorem}\label{the:proof-trees-qans-warded}
  Consider a database $D$, a set $\dep \in \class{WARD}$ of TGDs, a CQ
  $q(\bar x)$, and $\bar c \in \adom{D}^{|\bar x|}$. The following are
  equivalent:
\begin{enumerate}
\item $\bar c \in \cert{q}{D}{\dep}$.
\item There exists a proof tree $\mathcal{P}$ of $q$ w.r.t.~$\dep$
  with $\nwd{\cali{P}} \leq f_{\class{WARD}}(q,\dep)$ such that
  $\bar c \in \mathcal{P}(D)$.
\end{enumerate}
\end{theorem}

\noindent
\paragraph{A Proof Sketch.} Let us now provide some details on how Theorems~\ref{the:proof-trees-qans-pwl-warded} and~\ref{the:proof-trees-qans-warded} are shown.
For both theorems, (2) implies (1) readily follows from Theorem~\ref{the:proof-trees-qans}. We thus focus on the other direction.
The main ingredients of the proof can be described as follows:
\begin{itemize}
\item We introduce the auxiliary notion of chase tree, which can be seen as a concrete instantiation of a proof tree. It serves as an intermediate structure between proof trees and chase derivations, which allows us to use the chase as our underlying technical tool. Note that the notions of linearity and node-width can be naturally defined for chase trees.

\item We then show that, if the given tuple of constants $\bar c$ is a certain answer to the given CQ $q$ w.r.t.~the given database $D$ and (piece-wise linear) warded set $\dep$ of TGDs, then there exists a (linear) chase tree for the image of $q$ to $\chase{D}{\dep}$ such that its node-width respects the bounds given in the above theorems (Lemma~\ref{lem:existence-ct}).

\item We finally show that the existence of a (linear) chase tree for the image of $q$ to $\chase{D}{\dep}$ with node-width at most $m$ implies the existence of a (linear) proof tree $\mathcal{P}$ of $q$ w.r.t.~$\dep$ with node-width at most $m$ such that $\bar c \in \ca{P}(D)$ (Lemma~\ref{lem:from-ct-to-pt}).
\end{itemize}
Let us make the above description more formal. In order to introduce the notion of chase tree, we first need to recall the notion of chase graph, then introduce the notion of unravelling of the chase graph, and finally introduce the notions of unfolding and decomposition for sets of atoms in the unravelling of the chase graph.


%
%

Fix a chase sequence $\delta = (I_i \langle \sigma_i, h_i \rangle I_{i+1})_{i \geq 0}$ for a database $D$ under a set $\Sigma$ of TGDs. The \emph{chase graph} for $D$ and $\dep$ (w.r.t.~$\delta$) is a directed edge-labeled graph $\ca{G}^{D,\dep} = \tup{V,E,\lambda}$, where $V = \chase{D}{\Sigma}$, and an edge $(\alpha,\beta)$ labeled with $\tup{\sigma_k, h_k}$ belongs to $E$ iff $\alpha \in h_k(\body{\sigma_k})$ and $\beta \in I_{k+1} \setminus I_k$, for some $k \geq 0$. In other words, $\alpha$ has an edge to $\beta$ if $\beta$ is derived using $\alpha$, and if $\beta$ is new in the sense that it has not been derived before. Notice that $\ca{G}^{D,\dep}$ has no directed cycles.
Notice also that $\ca{G}^{D,\dep}$ depends on $\delta$ -- however, we can assume a fixed sequence $\delta$ since, as discussed in Section~\ref{sec:preliminaries}, every chase sequence is equally useful for our purposes.
%

We now discuss the notion of unravelling of the chase graph; due to space reasons, we keep this discussion informal. Given a set $\Theta \subseteq \chase{D}{\Sigma}$, the \emph{unraveling of $\ca{G}^{D,\dep}$ around $\Theta$} is a directed node- and edge-labeled forest $\ca{G}^{D,\dep}_\Theta$ that has a tree for each $\alpha \in \Theta$ whose branches are backward-paths in $\ca{G}$ from $\alpha$ to a database atom.
Intuitively, $\ca{G}^{D,\dep}_\Theta$ is a forest-like reorganization of the atoms of $\chase{D}{\Sigma}$ that are needed to derive $\Theta$. Due to its forest-like shape, it may contain multiple copies of atoms of $\chase{D}{\Sigma}$. The edges between nodes are labeled by pairs $\tup{\sigma,h}$ just like in $\ca{G}^{D,\dep}$, while the nodes are labeled by atoms and, importantly, the atoms along the paths in $\ca{G}^{D,\dep}$ may be duplicated and labeled nulls are given new names. We write $U(\ca{G}^{D,\dep},\Theta)$ for the set of all atoms that appear as labels in $\ca{G}^{D,\dep}_\Theta$, and $\succs{\alpha}{\sigma}{h}$ for the set of children of $\alpha$ whose incoming edge is labeled with $\tup{\sigma, h}$. It is important to say that there exists a homomorphism $h_\Theta$ that maps $\Theta$ to $U(\ca{G}^{D,\dep},\Theta)$.


Let us now introduce the notions of unfolding and decomposition.
For sets $\Gamma, \Gamma' \subseteq U(\ca{G}^{D,\dep}, \Theta)$, we say that
{\em $\Gamma'$ is an unfolding of $\Gamma$}, if there are $\alpha \in \Gamma$ and $\beta_1,\ldots,\beta_k \in U(\ca{G}^{D,\dep},\Theta)$ such that
\begin{enumerate}
\item $\succs{\alpha}{\sigma}{h} = \set{\beta_1,\ldots,\beta_k}$, for some
  $\sigma \in \Sigma$ and $h$,

\item for every null that occurs in $\alpha$, either it does not appear
  in $\Gamma \setminus \set{\alpha}$, or it appears in $\set{\beta_1,\ldots,\beta_k}$, and

\item $\Gamma' = (\Gamma \setminus \set{\alpha}) \cup
  \set{\beta_1,\ldots,\beta_k}$.
\end{enumerate}
Let $\Gamma \subseteq U(\ca{G}^{D,\dep},\Theta)$ be a non-empty set. A \emph{decomposition} of $\Gamma$ is a set $\{\Gamma_1,\ldots,\Gamma_n\}$, where $n \geq 1$, of non-empty subsets of $\Gamma$ such that (i) $\Gamma = \bigcup_{i \in [k]} \Gamma_i$, and (ii) $i \neq j$ implies that $\Gamma_i$ and $\Gamma_j$ do not share a labeled null.
We can now define the key notion of chase tree:

%


\begin{definition}[\textbf{Chase Tree}]
Consider a database $D$, a set $\dep$ of TGDs, and a set $\Theta \subseteq \chase{D}{\dep}$. A \emph{chase tree} for $\Gamma \subseteq U(\ca{G}^{D,\dep},\Theta)$ (w.r.t.~$\ca{G}^{D,\dep}_\Theta$) is a pair $\mathcal{C} = (T,\lambda)$, where $T = (V,E)$ is a finite rooted tree, and $\lambda$ a labeling function that assigns a subset of $U(\ca{G}^{D,\dep},\Theta)$ to each node of $T$, such that, for each $v \in V$:
\begin{enumerate}
\item If $v$ is the root node of $T$, then $\lambda(v) = \Gamma$.

\item If $v$ has only one child $u$, then $\lambda(u)$ is an unfolding of $\lambda(v)$.

\item If $v$ has $k > 1$ children $u_1,\ldots,u_k$, then $\{\lambda(u_1),\ldots,\lambda(u_k)\}$ is a decomposition of $\lambda(v)$.

\item If $v$ is a leaf node, then $\lambda(v) \subseteq D$.
\end{enumerate}
The \emph{node-width} of $\ca{C}$ is $\nwd{\ca{C}} \coloneqq \max_{v \in V} \{|\lambda(v)|\}$. Moreover, we say that $\ca{C}$ is \emph{linear} if, for each node $v \in V$, there exists at most one $u \in V$ such that $(v,u) \in E$ and $u$ is not a leaf. \hfill\markfull
\end{definition}

We can now state our auxiliary technical lemmas. In what follows, fix a database $D$, and a set $\dep$ of TGDs.


%

\begin{lemma}\label{lem:existence-ct}
Let $\Theta \subseteq \chase{D}{\dep}$ and $\Gamma \subseteq U(\ca{G}^{D,\dep},\Theta)$. Then:
\begin{enumerate}
\item If $\Sigma \in \class{WARD} \cap \class{PWL}$, then there exists
    a linear chase tree $\ca{C}$ for $\Gamma$ such that $\nwd{\ca{C}} \leq f_{\class{WARD} \cap \class{PWL}}(\Gamma, \Sigma)$.

\item If $\Sigma \in \class{WARD}$, then there exists a chase tree $\ca{C}$ for $\Gamma$ such that $\nwd{\ca{C}} \leq f_{\class{WARD}}(\Gamma, \Sigma)$.
\end{enumerate}
\end{lemma}

The next technical lemma exposes the connection between chase trees and proof
trees:

\begin{lemma}\label{lem:from-ct-to-pt}
Consider a set $\Theta \subseteq \chase{D}{\dep}$, and let $q'(\bar x)$ be a CQ and $\bar c$ a tuple of constants such that $h'(\atoms{q'}) \subseteq U(\ca{G}^{D,\dep},\Theta)$ and $h'(\ve{x}) = \ve{c}$, for some homomorphism $h'$.
If there is a (linear) chase tree $\ca{C}$ for $h'(\atoms{q'})$ with $\nwd{\ca{C}} \leq m$, then there is a (linear) proof tree $\ca{P}$ for $q'$ w.r.t.~$\Sigma$ such that $\nwd{\ca{P}} \leq m$ and $\ve{c} \in \ca{P}(D)$.
\end{lemma}


We now show Theorem~\ref{the:proof-trees-qans-pwl-warded}, while Theorem~\ref{the:proof-trees-qans-warded} can be shown analogously.
Consider a CQ $q(\bar x)$ and a tuple $\bar c \in \adom{D}^{|\bar x|}$ such that $\bar c \in \cert{q}{D}{\dep}$. We need to show that if $\dep \in \class{WARD} \cap \class{PWL}$, then there exists a linear proof tree $\ca{P}$ of $q$ w.r.t.~$\dep$ with $\nwd{\ca{P}} \leq f_{\class{WARD} \cap \class{PWL}}(q, \Sigma)$ such that $\bar c \in \ca{P}(D)$.
By hypothesis, there is a homomorphism $h$ such that $h(\atoms{q}) \subseteq \chase{D}{\dep}$ and $h(\bar x) = \bar c$. Let $\Theta_q$ be the set of atoms $h(\atoms{q})$.
Recall that there is a homomorphism $h_{\Theta_q}$ that maps $\Theta_q$ to $U(\ca{G}^{D,\dep},\Theta_q)$. Thus, the homomorphism $h' = h_{\Theta_q} \circ h$ is such that $h'(\atoms{q}) \subseteq U(\ca{G}^{D,\dep},\Theta_q)$ and $h'(\bar x) = \bar c$.
By Lemma~\ref{lem:existence-ct}, there exists a chase tree $\ca{C}$ for $h'(\atoms{q})$ with $\nwd{\ca{C}} \leq f_{\class{WARD} \cap \class{PWL}}(h'(\atoms{q}), \Sigma)$. By Lemma~\ref{lem:from-ct-to-pt}, there exists a linear proof tree $\ca{P}$ of $q$ w.r.t~$\dep$ with $\nwd{\ca{P}} \leq f_{\class{WARD} \cap \class{PWL}}(h'(\atoms{q}), \Sigma) \leq f_{\class{WARD} \cap \class{PWL}}(q, \Sigma)$ such that $\bar c \in \ca{P}(D)$, and the claim follows.

\subsection{Complexity Analysis}

\begin{algorithm}[t]
    \KwIn{$D$, $\dep \in \class{WARD} \cap \class{PWL}$, $q(\bar x)$, $\bar c \in \adom{D}^{|\bar x|}$}
    \KwOut{$\mathsf{Accept}$ if $\bar c \in \cert{q}{D}{\dep}$; otherwise, $\mathsf{Reject}$}
    \vspace{2mm}
    $p := Q \la \alpha_1,\ldots,\alpha_n$ with $\atoms{q(\bar c)} = \{\alpha_1,\ldots,\alpha_n\}$\\
    \Repeat{$\atoms{p} \subseteq D$}{
            \textbf{guess} $\mi{op} \in \{\mathsf{r},\mathsf{d},\mathsf{s}\}$\\
            \If{$\mi{op} = \mathsf{r}$}{\textbf{guess} a TGD $\sigma \in \dep$\\
            \eIf{$\mathsf{mgcu}(p,\sigma) = \emptyset$}{$\mathsf{Reject}$}{\textbf{guess} $U \in \mathsf{mgcu}(p,\sigma)$\\
            \eIf{$|p[\sigma,U]| > f_{\class{WARD} \cap \class{PWL}}(q,\dep)$}{$\mathsf{Reject}$}{$p' := p[\sigma,U]$}}}
            \If{$\mi{op} = \mathsf{d}$}{$p' := p[-D]$}
            \If{$\mi{op} = \mathsf{s}$}{\textbf{guess} $V \subseteq \var{p}$ and $\gamma: V \ra \adom{D}$\\
            $p' := \gamma(p)$}
            $p := p'$
    }
    \Return{$\mathsf{Accept}$}
\end{algorithm}

We now have all the tools for showing that CQ answering under piece-wise linear warded sets of TGDs is in \textsc{PSpace} in combined complexity, and in \textsc{NLogSpace} in data complexity, and also for re-establishing the complexity of warded sets of TGDs (see Proposition~\ref{the:warded-complexity}) in a more transparent way than the approach of~\cite{GoPi15}.

%
\medskip

\noindent
\paragraph{The Case of $\qans(\class{WARD} \cap \class{PWL})$.} Given a database $D$, a set $\dep \in \class{WARD} \cap \class{PWL}$ of TGDs, a CQ $q(\bar x)$, and a tuple $\bar c \in \adom{D}^{|\bar x|}$, by Theorem~\ref{the:proof-trees-qans-pwl-warded}, our problem boils down to checking whether there exists a linear proof tree $\mathcal{P}$ of $q$ w.r.t.~$\dep$ with $\nwd{\cali{P}} \leq f_{\class{WARD} \cap \class{PWL}}(q,\dep)$ such that $\bar c \in \mathcal{P}(D)$. This can be easily checked via a space-bounded algorithm that is trying to build such a proof tree in a level-by-level fashion. Essentially, the algorithm builds the $i$-th level from the $(i-1)$-th level of the proof tree by non-deterministically applying the operations introduced above, i.e., resolution, decomposition and specialization.

The algorithm is depicted in the box above. Here is a semi-formal description of it. The first step is to store in $p$ the Boolean CQ obtained after instantiating the output variables of $q$ with $\bar c$.
%
%
The rest of the algorithm is an iterative procedure that non-deterministically constructs $p'$ (the $i$-th level) from $p$ (the $(i-1)$-th level) until it reaches a level that is a subset of the database $D$. Notice that $p$ and $p'$ always hold one CQ since at each level of a linear proof tree only one node has a child, while all the other nodes are leaves, which essentially means that their atoms appear in the database $D$.
At each iteration, the algorithm constructs $p'$ from $p$ by applying resolution ($\mathsf{r}$), decomposition ($\mathsf{d}$), or specialization ($\mathsf{s}$):

\begin{description}
\item[{Resolution.}] It guesses a TGD $\sigma \in \dep$. If the set $\mathsf{mgcu}(p,\sigma)$, i.e., the set of all MGCUs of $p$ with $\sigma$, is empty, then rejects; otherwise, it guesses $U \in \mathsf{mgcu}(p,\sigma)$. If the size of the $\sigma$-resolvent of $p$ obtained via $U$, denoted $p[\sigma,U]$, does not exceed the bound given by Theorem~\ref{the:proof-trees-qans-pwl-warded}, then it assigns $p[\sigma,U]$ to $p'$; otherwise, it rejects.
    Recall that during a resolution step we need to rename variables in order to avoid undesirable clatter. However, we cannot blindly use new variables at each step since this will explode the space used by algorithm. Instead, we should reuse variables that have been lost due to their unification with an existentially quantified variable. A simple analysis shows that we only need polynomially many variables, while this polynomial depends only on $q$ and $\dep$.

\smallskip

\item[{Decomposition.}] It deletes from $p$ the atoms that occur in $D$, and it assigns the obtained CQ $p[-D]$ to $p'$. Notice that $p[-D]$ may be empty in case $\atoms{p} \subseteq D$. Essentially, the algorithm decomposes $p$ in such a way that the subquery of $p$ consisting of $\atoms{p} \cap D$ forms a child of $p$ that is a leaf, while the subquery consisting of $\atoms{p} \setminus D$ is the non-leaf child.

\smallskip

\item[{Specialization.}] It assigns to $p'$ a specialized version of $p$, where some variables are instantiated by constants of $\adom{D}$. Notice that the convention that output variables correspond to constants is implemented by directly instantiating them with actual constants from $\adom{D}$.
\end{description}

\noindent After constructing $p'$, the algorithm assigns it to $p$, and this ends one iteration.
%
If $\atoms{p} \subseteq D$, then a linear proof tree $\mathcal{P}$ such that $\bar c \in \mathcal{P}(D)$ has been found, and the algorithm accepts; otherwise, it proceeds with the next iteration.

It is easy to see that the algorithm uses polynomial space in general. Moreover, in case the set of TGDs and the CQ are fixed, the algorithm uses logarithmic space, which is the space needed for representing constantly many elements of $\adom{D}$; each element of $\adom{D}$ can be represented using logaritmically many bits. The desired upper bounds claimed in Theorem~\ref{the:warded-pwl-complexity} follow.

%
\medskip

\noindent
\paragraph{The Case of $\qans(\class{WARD})$.} The non-deterministic algorithm discussed above cannot be directly used for warded sets of TGDs since it is not enough to search for a linear proof tree as in the case of piece-wise linear warded sets of TGDs. However, by Theorem~\ref{the:proof-trees-qans-warded}, we can search for a proof tree that has bounded node-width. This allows us to devise a space-bounded algorithm, which is similar in spirit as the one presented above, with the crucial difference that it constructs in a level-by-level fashion the branches of the proof tree in parallel universal computations using alternation. Since this alternating algorithm uses polynomial space in general, and logarithmic space when the set of TGDs and the CQ are fixed, we immediately get an \textsc{ExpTime} upper bound in combined, and a \textsc{PTime} upper bound in data complexity. This confirms Proposition~\ref{the:warded-complexity} established in~\cite{GoPi15}. However, our new algorithm is significantly simpler than the one employed in~\cite{GoPi15}, while Theorem~\ref{the:proof-trees-qans-warded} reveals the main property of warded sets of TGDs that leads to the desirable complexity upper bounds.


\section{A Justified Combination}\label{sec:justification}

It is interesting to observe that the class of piece-wise linear warded sets of TGDs generalizes the class of {\em intensionally linear} sets of TGDs, denoted $\class{IL}$, where each TGD has at most one body atom whose predicate is intensional. Therefore, Theorem~\ref{the:warded-pwl-complexity} immediately implies that $\qans(\class{IL})$ is \textsc{PSpace}-complete in combined complexity, and \textsc{NLogSpace}-complete in data complexity. Notice that $\class{IL}$ generalizes linear Datalog, which is also \textsc{PSpace}-complete in combined complexity, and \textsc{NLogSpace}-complete in data complexity. Thus, we can extend linear Datalog by allowing existentially quantified variables in rule heads, which essentially leads to $\class{IL}$, without affecting the complexity of query answering.

At this point, one maybe tempted to think that the same holds for piece-wise linear Datalog, i.e., we can extend it with existentially quantified variables in rule heads, which leads to $\class{PWL}$, without affecting the complexity of query answering, that is, \textsc{PSpace}-complete in combined, and \textsc{NLogSpace}-complete in data complexity.
However, if this is the case, then wardedness becomes redundant since the formalism that we are looking for is the class of piece-wise linear sets of TGDs, without the wardedness condition. It turned out that this is not the case. To our surprise, the following holds:

\begin{theorem}\label{the:pwl-undecidable}
$\qans(\class{PWL})$ is undecidable in data complexity.
\end{theorem}

To show the above result we exploit an undecidable tiling problem~\cite{Boas97}. A {\em tiling system} is a tuple $\mathbb{T} = (T,L,R,H,V,a,b)$, where $T$ is a finite set of tiles, $L,R \subseteq T$ are special sets of left and right border tiles, respectively, with $L \cap R = \emptyset$, $H,V \subseteq T^2$ are the horizontal and vertical constraints, and $a,b$ are distinguished tiles of $T$ called the start and the finish tile, respectively.
A {\em tiling} for $\mathbb{T}$ is a function $f \colon [n] \times [m] \ra T$, for some $n,m > 0$, such that $f(1,1) = a$, $f(1,m) = b$, $f(1,i) \in L$ and $f(n,i) \in R$, for every $i \in [m]$, and $f$ respects the horizontal and vertical constraints. In other words, the first and the last rows of a tiling for $\mathbb{T}$ start with $a$ and $b$, respectively, while the leftmost and rightmost columns contain only tiles from $L$ and $R$, respectively.
We reduce from:

\medskip

\begin{center}
\fbox{\begin{tabular}{ll}
{\small PROBLEM} : & $\corr$
\\
{\small INPUT} : & A tiling system $\mathbb{T}$.
\\
{\small QUESTION} : &  Is there a tiling for $\mathbb{T}$?
\end{tabular}}
\end{center}

\medskip

Given a tiling system $\mathbb{T} = (T,L,R,H,V,a,b)$, the goal is to construct in polynomial time a database $D_{\mathbb{T}}$, a set of TGDs $\dep \in \class{PWL}$, and a Boolean CQ $q$, such that $\mathbb{T}$ has a tiling iff $() \in \cert{q}{D_{\mathbb{T}}}{\dep}$; $()$ is the empty tuple. Note that $\dep$ and $q$ should not depend on $\mathbb{T}$.

\medskip

\noindent
\paragraph{The Database $D_{\mathbb{T}}$.} It simply stores the tiling system $\mathbb{T}$:
\begin{eqnarray*}
\hspace{-5mm} &&  \{{\rm Tile}(t) \mid t \in T\}\ \cup\ \{{\rm Left}(t) \mid t \in L\}\ \cup \{{\rm Right}(t) \mid t \in R\}\\
\hspace{-5mm} &\cup&  \{H(t,t') \mid (t,t') \in H\}\ \cup\ \{V(t,t') \mid (t,t') \in V\}\\
\hspace{-5mm} &\cup& \{{\rm Start}(a),{\rm Finish}(b)\}.
\end{eqnarray*}

\smallskip

\noindent
\paragraph{The Set of TGDs $\dep$.} It is responsible for generating all the candidate tilings for $\mathbb{T}$, i.e., tilings without the condition $f(1,m) = b$, of arbitrary width and depth. Whether there exists a candidate tiling for $\mathbb{T}$ that satisfies the condition $f(1,m) = b$ it will be checked by the CQ $q$.
%
%
%
%
%
%
%
%
%
The set $\dep$ essentially implements the following idea: construct rows of size $\ell$ from rows of size $\ell-1$, for $\ell > 1$, that respect the horizontal constraints, and then construct all the candidate tilings by combining compatible rows, i.e., rows that respect the vertical constraints.
A row $r$ is encoded as an atom ${\rm Row}(p,c,s,e)$, where $p$ is the id of the row from which $r$ has been obtained, i.e., the previous one, $c$ is the id of $r$, i.e., the current one, $s$ is the starting tile of $r$, and $e$ is the ending tile of $r$. We write ${\rm Row}(c,c,s,s)$ for rows consisting of a single tile, which do not have a previous row (hence the id of the previous row coincides with the id of the current row), and the starting tile is the same as the ending tile.
The following two TGDs construct all the rows that respect the horizontal constraints:
\begin{align*}
{\rm Tile}(x)\ &\ra\ \exists z \, {\rm Row}(z,z,x,x),\\
{\rm Row}(\_,x,y,z), H(z,w)\ &\ra\ \exists u \,{\rm Row}(x,u,y,w).
\end{align*}
Analogously to Prolog, we write ``$\_$'' for a ``don't-care'' variable that occurs only once in the TGD.
The next set of TGDs constructs all the pairs of compatible rows, i.e., pairs of rows $(r_1,r_2)$ such that we can place $r_2$ below $r_1$ without violating the vertical constraints. This is done again inductively as follows:
\begin{eqnarray*}
&& \hspace{-7mm} {\rm Row}(x,x,y,y), {\rm Row}(x',x',y',y'),V(y,y')\ \ra\ {\rm Comp}(x,x'),\\
&& \hspace{-7mm} {\rm Row}(x,y,\_,z), {\rm Row}(x',y',\_,z'),\\
&& \hspace{25mm} {\rm Comp}(x,x'), V(z,z')\ \ra\ {\rm Comp}(y,y').
\end{eqnarray*}
We finally compute all the candidate tilings, together with their bottom-left tile, using the following two TGDs:
\begin{eqnarray*}
&& \hspace{-7mm} {\rm Row}(\_,x,y,z), {\rm Start}(y), {\rm Right}(z)\ \ra\ {\rm CTiling}(x,y),\\
&& \hspace{-7mm} {\rm CTiling}(x,\_), {\rm Row}(\_,y,z,w), {\rm Comp}(x,y),\\
&& \hspace{27mm} {\rm Left}(z), {\rm Right}(w)\ \ra\ {\rm CTiling}(y,z).
\end{eqnarray*}
This concludes the definition of $\dep$.

\medskip

\noindent
\paragraph{The Boolean CQ $q$.} Recall that $q$ is responsible for checking whether there exists a candidate tiling such that its bottom-left tile is $b$. This can be easily done via the query
\[
Q\ \leftarrow\ {\rm CTiling}(x,y), {\rm Finish}(y).
\]

By construction, the set $\dep$ of TGDs belongs to $\class{PWL}$. Moreover, there is a tiling for $\mathbb{T}$ iff $() \in \cert{q}{D_{\mathbb{T}}}{\dep}$, and Theorem~\ref{the:pwl-undecidable} follows.

\section{Expressive Power}\label{sec:expressiveness}

A class of TGDs naturally gives rise to a declarative database query language. More precisely, we consider queries of the form $(\dep,q)$, where $\dep$ is a set of TGDs, and $q$ a CQ over $\sch{\dep}$. The {\em extensional (database) schema} of $\dep$, denoted $\edb{\dep}$, is the set of extensional predicates of $\sch{\dep}$, i.e., the predicates that do not occur in the head of a TGD of $\dep$.
Given a query $Q = (\dep,q)$ and a database $D$ over $\edb{\dep}$, the {\em evaluation} of $Q$ over $D$, denoted $Q(D)$, is defined as $\cert{q}{D}{\dep}$.
We write $(\class{C},\class{CQ})$ for the query language consisting of all the queries $(\dep,q)$, where $\dep \in \class{C}$, and $q$ is a CQ. The evaluation problem for such a query language, dubbed $\eval(\class{C},\class{CQ})$, is defined in the usual way.
%
%
%
%
By definition, ${\bar c} \in Q(D)$ iff ${\bar c} \in \cert{q}{D}{\dep}$. Therefore, the complexity of $\eval(\class{C},\class{CQ})$ when $\class{C} = \class{WARD} \cap \class{PWL}$ and $\class{C} = \class{WARD}$ is immediately inherited from Theorem~\ref{the:warded-pwl-complexity} and Proposition~\ref{the:warded-complexity}, respectively:

\begin{theorem}
The following statements hold:
\begin{enumerate}
\item $\eval(\class{WARD} \cap \class{PWL},\class{CQ})$ is {\rm\textsc{PSpace}}-complete in combined complexity, and {\rm\textsc{NLogSpace}}-complete in data complexity.

\item $\eval(\class{WARD},\class{CQ})$ is {\rm\textsc{ExpTime}}-complete in combined complexity, and {\rm\textsc{PTime}}-complete in data complexity.
\end{enumerate}
\end{theorem}

The main goal of this section is to understand the relative expressive power of $(\class{WARD} \cap \class{PWL},\class{CQ})$ and $(\class{WARD},\class{CQ})$. To this end, we are going to adopt two different notions of expressive power namely the classical one, which we call combined expressive power since it considers the set of TGDs and the CQ as one composite query, and the program expressive power, which aims at the decoupling of the set of TGDs from the actual CQ. We proceed with the details starting with the combined expressive power.

\subsection{Combined Expressive Power}

Consider a query $Q = (\dep,q)$, where $\dep$ is a set of TGDs and $q(\bar x)$ a CQ over $\sch{\dep}$. The {\em expressive power} of $Q$, denoted $\ep{Q}$, is the set of pairs $(D,\bar c)$, where $D$ is a database over $\edb{\dep}$, and $\bar c \in \adom{D}^{|\bar x|}$, such that $\bar c \in Q(D)$.
The {\em combined expressive power} of a query language $(\class{C},\class{CQ})$, where $\class{C}$ is a class of TGDs, is defined as the set
\[
\cep{\class{C},\class{CQ}}\ =\ \{\ep{Q} \mid Q \in (\class{C},\class{CQ})\}.
\]
Given two query languages $\class{Q}_1, \class{Q}_2$, we say that $\class{Q}_2$ is {\em more expressive (w.r.t.~the combined expressive power)} than $\class{Q}_1$, written $\class{Q}_1 \leq_{\mathsf{cep}} \class{Q}_2$, if $\cep{\class{Q}_1} \subseteq\cep{\class{Q}_2}$.
Moreover, we say that $\class{Q}_1$ and $\class{Q}_2$ are {\em equally expressive (w.r.t.~the combined expressive power)}, written $\class{Q}_1 =_{\mathsf{cep}} \class{Q}_2$, if $\class{Q}_1 \leq_{\mathsf{cep}} \class{Q}_2$ and $\class{Q}_2 \leq_{\mathsf{cep}} \class{Q}_1$.

The next easy lemma states that $\class{Q}_1 =_{\mathsf{cep}} \class{Q}_2$ is equivalent to say that every query of $\class{Q}_1$ can be equivalently rewritten as a query of $\class{Q}_2$, and vice versa.
Given two query languages $\class{Q}_1$ and $\class{Q}_2$, we write $\class{Q}_1 \preceq \class{Q}_2$ if, for every $Q = (\dep,q) \in \class{Q}_1$, there exists $Q' = (\dep',q') \in \class{Q}_2$ such that, for every $D$ over $\edb{\dep} \cap \edb{\dep'}$, $Q(D) = Q'(D)$.
%

\begin{lemma}\label{lem:cep}
Consider two query languages $\class{Q}_1$ and $\class{Q}_2$. It holds that $\class{Q}_1 \leq_{\mathsf{cep}} \class{Q}_2$ iff $\class{Q}_1 \preceq \class{Q}_2$.
\end{lemma}



We are now ready to state the main result of this section, which reveals the expressiveness of $(\class{WARD} \cap \class{PWL},\class{CQ})$ and $(\class{WARD},\class{CQ})$ relative to Datalog.
Let us clarify that a Datalog query is essentially a pair $(\dep,q)$, where $\dep$ is a Datalog program, or a set of {\em full} TGDs, i.e., TGDs without existentially quantified variables, that have only one head atom, and $q$ a CQ. We write $\class{FULL}_1$ for the above class of TGDs. In other words, piece-wise linear Datalog, denoted $\class{PWL{\text{-}}DATALOG}$, is the language $(\class{FULL}_1 \cap \class{PWL},\class{CQ})$, while Datalog, denoted $\class{DATALOG}$, is the language $(\class{FULL}_1,\class{CQ})$, and thus we can refer to their combined expressive power.

\begin{theorem}\label{the:cep}
The following statements hold:
\begin{enumerate}
\item $\class{PWL{\text{-}}DATALOG}\ =_{\mathsf{cep}}\ (\class{WARD} \cap \class{PWL},\class{CQ})$.

\item $\class{DATALOG}\ =_{\mathsf{cep}}\ (\class{WARD},\class{CQ})$.
\end{enumerate}
\end{theorem}

Let us explain how (1) is shown; the proof for (2) is similar. We need to show that: (i) $\class{PWL{\text{-}}DATALOG} \leq_{\mathsf{cep}} (\class{WARD} \cap \class{PWL},\class{CQ})$, and (ii) $(\class{WARD} \cap \class{PWL},\class{CQ}) \leq_{\mathsf{cep}} \class{PWL{\text{-}}DATALOG}$.
%
%
By definition, $\class{FULL}_1 \cap \class{PWL} \subseteq \class{WARD} \cap \class{PWL}$. Thus, $(\class{FULL}_1 \cap \class{PWL},\class{CQ}) \preceq (\class{WARD} \cap \class{PWL},\class{CQ})$, which, together with Lemma~\ref{lem:cep}, implies (a).
For showing (b), by Lemma~\ref{lem:cep}, it suffices to show that:

\begin{lemma}\label{lem:ward-pwl-to-dat-cep}
$(\class{WARD} \cap \class{PWL},\class{CQ})\ \preceq\ \class{PWL{\text{-}}DATALOG}$.
\end{lemma}

The key idea underlying the above lemma is to convert a linear proof tree $\ca{P}$ of a CQ $q(\bar x)$ w.r.t.~a set $\dep \in \class{WARD} \cap \class{PWL}$ of TGDs into a piece-wise linear Datalog query $Q = (\dep',q'(\bar x))$ such that, for every database $D$ over $\edb{\dep}$, $\ca{P}(D) = Q(D)$. Roughly, each node of $\ca{P}$ together with its children, is converted into a full TGD that is added to $\dep'$. Assume that the node $v$  has the children $u_1,\ldots,u_k$ in $\ca{P}$, where $v$ is labeled by $p_0(\bar x_0)$ and, for $i \in [k]$, $u_i$ is labeled by the CQ $p_i(\bar x_i)$ with $\bar x \subseteq \bar x_i$. In this case, we add to $\dep'$ the full TGD
\[
C_{[p_1]}(\bar x_1), \ldots, C_{[p_k]}(\bar x_k)\ \ra\ C_{[p_0]}(\bar x_0),
\]
where $C_{[p_i]}$ is a predicate that corresponds to the CQ $p_i$, while $[p_i]$ refers to a {\em canonical renaming} of $p_i$. The intention underlying such a canonical renaming is the following: if $p_i$ and $p_j$ are the same up to variable renaming, then $[p_i] = [p_j]$.
We also add to $\dep'$ a full TGD
\[
R(x_1,\ldots,x_n)\ \ra\ C_{[p_R]}(x_1,\ldots,x_n)
\]
for each $n$-ary predicate $R \in \edb{\dep}$, where $p_R(x_1,\ldots,x_n)$ is the atomic query consisting of the atom $R(x_1,\ldots,x_n)$.
Since in $\ca{P}$ we may have several CQs that are the same up to variables renaming, the set $\dep'$ is recursive, but due to the linearity of $\ca{P}$, the employed recursion is piece-wise linear, i.e., $\dep' \in \class{FULL}_1 \cap \class{PWL}$.
The CQ $q'(\bar x)$ is simply the atomic query $C_{[q]}(\bar x)$. It should not be difficult to see that indeed $\ca{P}(D) = Q(D)$, for every database $D$ over $\edb{D}$.

Having the above transformation of a linear proof tree into a piece-wise linear Datalog query in place, we can easily rewrite every query $Q = (\dep,q) \in (\class{WARD} \cap \class{PWL},\class{CQ})$ into an equivalent query that falls in $\class{PWL{\text{-}}DATALOG}$. We exhaustively convert each linear proof tree $\ca{P}$ of $q$ w.r.t.~$\dep$ such that $\nwd{\ca{P}} \leq f_{\class{WARD} \cap \class{PWL}}(q,\dep)$ into a piece-wise linear Datalog query $Q_{\ca{P}}$, and then we take the union of all those queries. 
Since we consider the canonical renaming of the CQs occurring in a proof tree, and since the size of those CQs is bounded by $f_{\class{WARD} \cap \class{PWL}}(q,\dep)$, we immediately conclude that we need to explore finitely many CQs. Thus, the above iterative procedure will eventually terminate and construct a finite piece-wise linear Datalog query that is equivalent to $Q$, as needed.

%
%


\subsection{Program Expressive Power}

The {\em expressive power} of a set $\dep$ of TGDs, denoted $\ep{\dep}$, is the set of triples $(D,q(\bar x),\bar c)$, where $D$ is a database over $\edb{\dep}$, $q(\bar x)$ is a CQ over $\sch{\dep}$, and $\bar c \in \adom{D}^{|\bar x|}$, such that $\bar c \in \cert{q}{D}{\dep}$.
The {\em program expressive power} of a query language $(\class{C},\class{CQ})$, where $\class{C}$ is a class of TGDs, is defined as the set
\[
\pep{\class{C},\class{CQ}}\ =\ \{\ep{\dep} \mid \dep \in \class{C}\}.
\]
Given two query languages $\class{Q}_1, \class{Q}_2$, we say that $\class{Q}_2$ is {\em more expressive (w.r.t.~program expressive power)} than $\class{Q}_1$, written $\class{Q}_1 \leq_{\mathsf{pep}} \class{Q}_2$, if $\pep{\class{Q}_1} \subseteq \pep{\class{Q}_2}$.
Moreover, we say that $\class{Q}_2$ is {\em strictly more expressive (w.r.t.~the program expressive power)} that $\class{Q}_2$, written $\class{Q}_1 <_{\mathsf{pep}} \class{Q}_2$, if $\class{Q}_1 \leq_{\mathsf{pep}} \class{Q}_2$ and $\class{Q}_2 \not\leq_{\mathsf{pep}} \class{Q}_1$.

Let us now establish a useful lemma, analogous to Lemma~\ref{lem:cep}, which reveals the essence of the program expressive power. For brevity, given two classes of TGDs $\class{C}_1$ and $\class{C}_2$, we write $\class{C}_1 \preceq \class{C}_2$ if, for every $\dep \in \class{C}_1$, there exists $\dep' \in \class{C}_2$ such that, for every $D$ over $\edb{\dep} \cap \edb{\dep'}$, and CQ $q$ over $\sch{\dep} \cap \sch{\dep'}$, $Q(D) = Q'(D)$, where $Q = (\dep,q)$ and $Q' = (\dep',q)$. The following holds:

\begin{lemma}\label{lem:pep}
Consider two query languages $\class{Q}_1 = (\class{C}_1,\class{CQ})$ and $\class{Q}_2 = (\class{C}_2,\class{CQ})$. Then, $\class{Q}_1 \leq_{\mathsf{pep}} \class{Q}_2$ iff $\class{C}_1 \preceq \class{C}_2$.
\end{lemma}


We are now ready to study the expressiveness (w.r.t.~the program expressive power) of $(\class{WARD} \cap \class{PWL},\class{CQ})$ and $(\class{WARD},\class{CQ})$ relative to Datalog. In particular, we show that:

\begin{theorem}\label{the:pep}
The following statements hold:
\begin{enumerate}
\item $\class{PWL{\text{-}}DATALOG} <_{\mathsf{pep}} (\class{WARD} \cap \class{PWL},\class{CQ})$.

\item $\class{DATALOG} <_{\mathsf{pep}} (\class{WARD},\class{CQ})$.
\end{enumerate}
\end{theorem}

Let us explain how (1) is shown; the proof for (2) is similar. We need to show that: (i) $\class{PWL{\text{\text{-}}}DATALOG} \leq_{\mathsf{pep}} (\class{WARD} \cap \class{PWL},\class{CQ})$, and (ii) $(\class{WARD} \cap \class{PWL},\class{CQ}) \not\leq_{\mathsf{pep}} \class{PWL{\text{-}}DATALOG}$.
Since, by definition, $\class{FULL}_1 \cap \class{PWL} \subseteq \class{WARD} \cap \class{PWL}$, we immediately get that $\class{FULL}_1 \cap \class{PWL} \preceq \class{WARD} \cap \class{PWL}$, and thus, by Lemma~\ref{lem:pep}, (a) follows.
For showing (b), by Lemma~\ref{lem:pep}, it suffices to show that:

\begin{lemma}\label{lem:ward-pwl-to-dat-pep}
$\class{WARD} \cap \class{PWL} \npreceq \class{FULL}_1 \cap \class{PWL}$.
\end{lemma}

By contradiction, assume the opposite. We define the set of TGDs $\dep = \{P(x) \ra \exists y \, R(x,y)\}$, the database $D = \{P(c)\}$, and the CQs
$q_1 = Q \la R(x,y)$ and $q_2 = Q \la R(x,y),P(y)$.
By hypothesis, there exists $\dep' \in \class{FULL}_1 \cap \class{PWL}$ such that $Q_1(D) = Q'_1(D)$ and $Q_2(D) = Q'_2(D)$, where $Q_i = (\dep,q_i)$ and $Q'_i = (\dep',q_i)$, for $i \in \{1,2\}$.
Clearly, $Q_1(D) \neq \emptyset$ and $Q_2(D) = \emptyset$, which implies that $Q'_1(D) \neq \emptyset$ and $Q'_2(D) = \emptyset$. However, it is easy to see that $Q'_1(D) \neq \emptyset$ implies $Q'_2(D) \neq \emptyset$, which is a contradiction, and the claim follows.


\section{Implementation and Future Work}\label{sec:conclusions}
%

%
The Vadalog system is currently optimized for piece-wise linear warded sets of TGDs in three ways. Here is brief description in order to give a glimpse which parts of the implementation are affected:

\begin{enumerate}[itemindent=0pt,labelwidth=1em,labelindent=0em,leftmargin=*]
\item The first one is related to the core of the system, namely the way that existential quantifiers interact with recursion. For this purpose, the system builds guide structures, the \emph{linear forest}, \emph{warded forest} and \emph{lifted linear forest}; for details see~\cite{BeSG18}. These structures are essential for {\em aggressive termination control}, i.e., terminating recursion as early as possible, and they are affected by the linearity/non-linearity of the TGDs. When the system operates on piece-wise linear warded sets of TGDs, these structures become (by design) more effective at terminating recursion earlier, having a significant effect on the memory footprint.

\item The optimizer detects and uses piece-wise linearity for the purpose of join ordering. In particular, many join algorithms are optimized towards having the recursive predicate as the first (or last) operand. Piece-wise linearity allows to distinguish {\em the} body atom of a TGD that is mutually recursive with a head atom, which allows the optimizer to be biased towards selecting this special atom as the first (or last) operand of the join.

\item The third way is related to the architecture of the system.  The Vadalog system builds from the plan constructed by the optimizer a network of operator nodes. This allows streaming of data through such a system. Differently from most database systems, recursion and existential quantification are directly considered within this network of nodes. This includes considering the guide structures mentioned above at most nodes, to allow for aggressive termination control. 
    The stratification induced by piece-wise linearity affects this network. In particular, the system may decide to insert materialization nodes at the boundaries of these strata, materializing intermediate results. Notice that this third point is a trade-off, as it actually raises memory footprint, but in turn can provide a speed-up.
\end{enumerate}


%
%
%

Here are some promising directions that we are planning to study in our future research:
\begin{enumerate}[itemindent=0pt,labelwidth=1em,labelindent=0em,leftmargin=*]
\item As said in Section~\ref{sec:introduction}, \textsc{NLogSpace} is contained in the class \textsc{NC}$_2$ of highly parallelizable problems. This means that reasoning under piece-wise linear warded sets of TGDs is principally parallelizable, unlike warded sets of TGDs. We plan to exploit this for the parallel execution of reasoning tasks in both multi-core settings and in the map-reduce model. In fact, we are currently in the process of implementing a multi-core implementation for piece-wise linear warded sets of TGDs. Our preliminary results are promising, giving evidence that the parallelization that is theoretically promised is also practically achievable.

\item Reasoning with piece-wise linear warded sets of TGDs is \textsc{LogSpace}-equivalent to reachability in directed graphs. Reachability in very large graphs has been well-studied and many algorithms and heuristics have been designed that work well in practice~\cite{CHKZ03,GBSW10,JiXRW08,King99,YiCZ10}. We are confident that several of these algorithms can be adapted for our purposes.
    
\item Finally, reachability in directed graphs is known to be in the {\em dynamic} parallel complexity class \textsc{Dyn-FO}~\cite{DKMSZ15,PaIm97}. This means that by maintaining suitable auxiliary data structures when updating a graph, reachability testing can actually be done in FO, and thus in SQL. We plan to analyze whether reasoning under piece-wise linear warded sets of TGDs, or relevant subclasses thereof, can be shown to be in \textsc{Dyn-FO}.
\end{enumerate}


\renewcommand{\baselinestretch}{.95}

\bibliographystyle{ACM-Reference-Format}
\bibliography{references}



\end{document}